\documentclass[12pt,a4paper]{article}
\usepackage{amsmath,amssymb,graphicx,bm,color}
\usepackage{float}
\numberwithin{equation}{section}
\allowdisplaybreaks[3]

\newcommand{\cI}{\mathcal{I}}
\newcommand{\cN}{\mathcal{N}}

\newcommand{\bsz}{\Bbb Z}
\newcommand{\bsr}{\Bbb R}

\newcommand{\dket}[1]{{\left.\left|#1\right\rangle\right\rangle}}
\newcommand{\dbra}[1]{{\left\langle\left\langle#1\right|\right.}}

\newcommand{\al}{\alpha}
\newcommand{\ep}{\epsilon}

\newcommand{\msc}[1]{\mbox{\scriptsize #1}}
\newcommand{\dsp}{\displaystyle}
\newcommand{\scs}[1]{{\scriptstyle #1}}

\newcommand{\br}{\Bbb R}
\newcommand{\bz}{\Bbb Z}

\renewcommand{\-}{{\bf -1}}

\newcommand{\cJ}{{\cal J}}
\newcommand{\cT}{{\cal T}}

\newcommand{\cP}{{\cal P}}

\newcommand{\cB}{{\cal B}}

\newcommand{\tL}{\tilde{L}}

\newcommand{\tcI}{\widetilde{\cal I}}

\renewcommand{\th}{{\theta}}

\newcommand{\tr}{\mbox{Tr}}

\renewcommand{\Re}{\mbox{Re}}

\newcommand{\nn}{\nonumber\\}

\newcommand{\NS}{\mbox{NS}}
\newcommand{\tNS}{\widetilde{\mbox{NS}}}
\newcommand{\R}{\mbox{R}}
\newcommand{\tR}{\widetilde{\mbox{R}}}
\newcommand{\sNS}{\msc{NS}}
\newcommand{\stNS}{\widetilde{\msc{NS}}}
\newcommand{\sR}{\msc{R}}

\newcommand{\any}{{}^{\forall}}

\newcommand{\bigbox}[2]{{\scs{#1}} \hspace{0.3mm}  \raise-1mm\hbox{$ \underset{#2} {\scalebox{1.6}{\mbox{$\Box$}}} $}}

\newcommand {\eqn}[1]{(\ref{#1})}

\makeatletter
\@addtoreset{equation}{section}

\makeatother

\begin{document}
%\maketitle

%%%%%%%%%%%%%%%%%%%%%%%%%%%%%%%%%%%%%%%%%%%%%%%%%%%%%%%%%%%%%%%%%%%%%%%%%%%%
%			paper

%%% Title page %%%%%
\begin{titlepage}
		
\renewcommand{\thefootnote}{\fnsymbol{footnote}}
\begin{flushright}
\begin{tabular}{l}
%0000-00-00\\
\today %This should be commented out.
\end{tabular}
\end{flushright}
		
\vfill
\begin{center}
			
% \vskip 2.5 truecm
			
\noindent{\Large \textbf{Non-BPS Fractional Branes}}
			
\medskip
			
\noindent{\Large \textbf{with Bose-Fermi Cancellation}}

\medskip
			
\noindent{\Large \textbf{in Asymmetric Orbifolds}} 
			
\vspace{1.5cm}

\noindent{Yuji Sugawara\footnote{E-mail: ysugawa@se.ritsumei.ac.jp} and Takahiro Uetoko\footnote{E-mail: rp0019fr@ed.ritsumei.ac.jp}}

\bigskip
			
\vskip .6 truecm
			
\centerline{\it Department of Physical Sciences, College of Science and Engineering,} 
\centerline{\it Ritsumeikan University, Shiga 525-8577, Japan}

\end{center}
		
\vfill
\vskip 0.5 truecm

\begin{abstract}

We study non-BPS D-branes in the type II string vacua with chiral space-time SUSY constructed based on the asymmetric orbifolds of 
 $\mbox{K3} \cong T^4/\bz_2$ as a succeeding work of \cite{SSU}.
We especially focus on the fractional D-branes that contains the contributions from the 
twisted sector of the $\bz_2$-orbifolding.

We show that the cylinder partition functions for these fractional branes do not vanish,  
as in the cases of ordinary non-BPS D-branes.
This aspect is in a sharp contrast with the bulk-type branes studied in \cite{SSU}.
%%%
We then discuss the extensions of models by including the discrete torsion
{\em depending on the spin structures\/}, 
and investigate whether we obtain the vanishing self-overlaps associated to the fractional branes.
The existence/absence of bose-fermi cancellations both in the closed and open string spectra 
as well as the massless spectra in the twisted sectors crucially depend on the discrete torsion. 
We find that some choices of the torsion indeed realize the vanishing self-overlaps for the fractional 
branes, with keeping the vanishing torus partition function intact.

\end{abstract}
\vfill
\vskip 0.5 truecm
		
\setcounter{footnote}{0}
\renewcommand{\thefootnote}{\arabic{footnote}}
\end{titlepage}
	
\newpage
	
\tableofcontents
%%%%%%%%%%%%%%%%%%%%%%%%%%%%%%%%%%%%%%%%%%%%%%%%%%%%%%%%%%%%%%%%%%%%%%

%%%%%%%%%%%%%%%%%%%%%%%%%%%%%%%%%%%%%%%%%%%%%%%%%%%%%%%%
%%%%%%%%%%%%%%%%%%%%%%%%%%%%%%%%%%%%%%%%%%%%%%%%%%%%%%%%

\section{Introduction}

Orbifold theories are important and interesting subjects in string theory, yielding various models of string vacua. 
Besides the symmetric ones, we can work with the asymmetric orbifolds \cite{Narain:1986qm}, 
in which the orbifold groups act asymmetrically on the left- and right-moving degrees of freedom, 
and they provide  non-geometric string vacua that are well-controlled.  
%%% 
%Furthermore, an interesting feature of orbifolds is 
%that they have some degrees of freedom of the phase factor known as the `discrete torsion'\cite{Vafa:1986wx,Vafa:1994rv,Gaberdiel:2000fe}.
%%%
One of interesting  studies  of asymmetric orbifolds would be the construction of non-SUSY vacua with perturbatively vanishing cosmological constant given {\em e.g.} in
\cite{Kachru1,Kachru2,Kachru3,Harvey,Shiu-Tye,Blumenhagen:1998uf,Angelantonj:1999gm,Antoniadis,Aoki:2003sy},
and  more recently  in \cite{SSW,SWada} based on simpler cyclic orbifold groups.
%%%%

Another interesting possibility to achieve non-SUSY vacua with very small cosmological constant would be realized 
by taking D-branes into account. 
Even though the space-time SUSY is preserved in the closed string sector, it would be broken by the effect of the non-BPS `D-brane instantons' 
(Euclidean D-branes wrapping around internal cycles), if we have sufficiently generic configurations of non-BPS D-branes.
%%%%
If we only have the vanishing $O(g_s^0)$-contributions to the cosmological constant 
originating from the open strings, that is, the cylinder partition functions associated to relevant non-BPS D-branes,  
%as well as the bulk ones,  
we would be left with a non-perturbative cosmological constant induced by the instanton effect.
%%%
%in other words, 
%the contributions coming from the world-sheets with the external legs sourced 
%by those non-BPS D-branes.
%%%
This should be exponentially suppressed as long as the string coupling $g_s$ is small enough.
Such a possibility in a type II theory has indeed been mentioned in \cite{Harvey} 
in order to explain the non-perturbative mismatch of the vacuum energy with that of its heterotic dual.
%%%
Closely related studies on non-SUSY vacua in heterotic string have 
also been given {\em e.g.} in
\cite{Blaszczyk:2014qoa,Angelantonj:2014dia,Faraggi:2014eoa,Abel:2015oxa,
Kounnas:2015yrc,Abel:2017rch}.
%%%%

%%%%%%%%%%%%%%%%%%%%%%%%%%%%%%%%%%%%%%%%%%%%%%%%%%%%%%%%%%%%%%%%%%%%%%%%%%%%%%%%%%%%%%%%%%%%%%%

In the recent paper \cite{SSU}, with this motivation, we studied non-BPS D-branes in some
%non-geometric 
type II string vacua with {\em chiral\/} space-time SUSY easily constructed 
by the asymmetric orbifolding.
%Especially, we have focused on Gepner construction for $K3$. 
%A crucial point of our idea is the chiral reflection $\sigma$ which eliminates all the left-moving supercharge. 
What is a simple but crucial fact is that, in these chiral SUSY vacua,
any boundary states $\dket{\cdots}$, $\dbra{\cdots}$ cannot satisfy the BPS-equation written as 
\begin{align}
	\left[ Q^\alpha + M^\alpha_{\;\;\beta} \tilde{Q}^\beta \right] \dket{\cB} = 0,
\label{BPSeq}
\end{align}
%
%where $Q^\alpha \, (\tilde{Q}^\beta)$ denotes the left(right)-moving space-time supercharges and %$M^\alpha_{\;\;\beta}$ are some c-number coefficients. 
just due to the lack of 
{\em e.g.} left-moving unbroken  supercharges $Q^{\alpha}$. 
%%%%
This means that any D-branes are non-BPS in these vacua. 
Nevertheless, as we demonstrated in \cite{SSU}, 
 the `self-overlaps'\footnote{In this paper, we call the cylinder partition function  
of which both ends are attached to the common BPS D-brane the `self-overlap'.} 
for these branes could vanish generically due to the bose-fermi cancellation. 
%%%%
%Various studies on non-BPS configurations of D-branes that however realize the bose-fermi cancellation 
%of open string excitations include {\em e.g.} \cite{Blumenhagen:1998uf,GSen}.
%%%
For instance, in the simplest example of the chiral orbifold of $T^4$ by the involution 
$\sigma \equiv (-1)^{F_L} \otimes (\-_R)^{\otimes 4} $, 
any non-BPS branes expressed as 
\begin{align}
\dket{\cB} = \sqrt{2} \cP_{\sigma} \dket{B}_{T^4},
\label{bulk brane 0}
\end{align}
possess the expected property, 
where $\cP_{\sigma}$ denotes the projection for the asymmetric orbifolding by $\sigma$ and 
$\dket{B}_{T^4}$ is an arbitrary boundary states describing a BPS brane in $T^4$.
We note that  these D-brane configurations are fairly generic 
in the theory $T^4/\sigma$.
%%%%%%%%%%%%%%%%%%%%%%%%%%%%%%%%%%%%%%%%%%%%%%%%%%%%%%%%%%%%%%%%%%%%%%%
%Recall that the twisted sector of $\sigma$ cannot contribute to 
%boundary states due to the asymmetric action of $\sigma$. Namely, we do not have 
%the 'fractional branes' in this vacuum.  
%%%
% cite the paper 'fractional branes in asymmetric orbifolds'
Preceding studies on non-BPS configurations of D-branes that however realize the bose-fermi cancellation 
of open string excitations include {\em e.g.} \cite{Blumenhagen:1998uf,GSen}.

%%%%%%%%%%%%%%%%%%%%%%%%%%%%%%%%%%%%%%%%%%%%%%%%%%%%%%%%%%%%%%%%%%%%%%%%%%%%%%%%%%%%%%%%%%%%%%
%%%%%%%%%%%%%%%%%%%%%%%%%%%%%%%%%%%%%%%%%%%%%%%%%%%%%%%%%%%%%%%%%%%%%%%%%%%%%%%%%%%%%%%%%%%%%%
%%%%%%%%%%%%%%%%%%%%%%%%%%%%%%%%%%%%%%%%%%%%%%%%%%%%%%%%%%%%%%%%%%%%%%%%%%%%%%%%%%%%%%%%%%%%%%

Now, in the present paper, we shall proceed to analyze  the non-BPS D-branes 
in the cases of asymmetric orbifolds of $\mbox{K3} \cong T^4/\bz_2$, 
where the $\bz_2$-orbifolding is defined by the symmetric reflection 
$\cI_4 \equiv (\-_L)^{\otimes 4} \otimes (\-_R)^{\otimes 4}$.
%%%
For the bulk-type branes 
$\dket{\cB} = \sqrt{2} \cP_{\sigma} \cP_{\cI_4} \dket{B}_{T^4}$
such as \eqn{bulk brane 0}, we just arrive at the same conclusion, since 
the argument given in \cite{SSU} is generic enough.
However, we can still accommodate the fractional branes in these vacua, 
that is, the boundary states including the contributions from the $\cI_4$-twisted sectors. 
The main aim of this paper is to examine whether we can also achieve the vanishing self-overlaps 
for these fractional branes.

%%%%%%%%%%%%%%%%%%%%%%%%%%%%%%%%%%%%%%%%%%%%%%%%%%%%%%%%%%%%%%%%%%%%%%
%%%%%%%%%%%%%%%%%%%%%%%%%%%%%%%%%%%%%%%%%%%%%%%%%%%%%%%%%%%%%%%%%%%%%%

The organization of this paper is given as follows:
%%%
In section 2, we first describe the boundary states  defined on the asymmetric orbifold 
$
\dsp 
\mbox{K3}/\sigma \equiv \left[T^4/\cI_4\right] / \sigma,
$
and analyze the cylinder amplitudes. 
%%%
We especially show that 
the self-overlaps associated to the fractional D-branes do not vanish generically
contrary to the bulk-type branes.
%not including the $\cI_4$-twisted sector.  
%as expected for ordinary non-BPS configurations of D-branes. 

Therefore, in section 3,  we shall discuss the extensions of models so as to include 
the discrete torsion \cite{Vafa:1986wx,Vafa:1994rv,Gaberdiel:2000fe} {\em depending on the spin structure\/}.
We can no longer regard the relevant conformal model  as a simple orbifold;
$
\dsp 
\mbox{K3}/\sigma \equiv \left[T^4/\cI_4\right] / \sigma,
$
and physical aspects non-trivially depend on the choice of discrete torsion. 
We classify the possible string vacua and investigate whether we achieve the vanishing  partition functions for
the world-sheet of torus as well as cylinder. 
We find out some choices of discrete torsion indeed realize the expected properties, 
namely, both of the torus partition function and the cylinder amplitudes for the fractional branes
vanish exactly. We also investigate extra massless excitations emerging in the twisted sectors 
both in the closed and open string Hilbert spaces of each model.

Furthermore, we consider the `stable non-BPS branes' in the K3-background given in \cite{GSen,BargG}, of which NSNS/RR components of the boundary states 
lie in the untwisted/twisted sectors.
So, we discuss what happens for the fractional branes of this type, after performing the $\sigma$-orbifolding.  
We find that relevant cylinder partition functions could vanish if choosing suitably the zero-mode parts of boundary states, 
in the similar but different way in comparison with \cite{GSen}.
This feature curiously depends on the discrete torsion. 
We argue that, in some cases,  the requirement for vanishing cylinder amplitudes is not compatible with the unitary open string spectra.

In section 4, we summarize the results in this paper and yield a brief discussion.

%%%%%%%%%%%%%%%%%%%%%%%%%%%%%%%%%%%%%%%%%%%%%%%%%%%%%%%%%%%%%%%%%%%%%%%%%%%%%%%%%%%%%%%%%%%%%%%
%%%%%%%%%%%%%%%%%%%%%%%%%%%%%%%%%%%%%%%%%%%%%%%%%%%%%%%%%%%%%%%%%%%%%%%%%%%%%%%%%%%%%%%%%%%%%%%
%%%%%%%%%%%%%%%%%%%%%%%%%%%%%%%%%%%%%%%%%%%%%%%%%%%%%%%%%%%%%%%%%%%%%%%%%%%%%%%%%%%%%%%%%%%%%%%

~

\section{Asymmetric orbifold without discrete torsion}

In this section we shall study the asymmetric orbifold without discrete torsion. 
%Let us assume the torus is along the $X^{6,\dots,9}$-directions at 
We start with the type II string vacuum compactified on the 4-dim. torus along the $X^{6,\dots,9}$-directions at the symmetry enhancement point with $\widehat{SO}(8)_1$. 
The total orbifold group is $\bz_2 \times \bz_2$ generated by $\cI_4:X^i\rightarrow-X^i\;(i=6,\dots,9)$
and the `chiral reflection'\footnote
   {Throughout this paper, we assume that $\sigma^2 = {\bf 1}$ for the untwisted sector with all the spin structures, although it is not necessarily satisfied 
for the Ramond sector as demonstrated in \cite{SSW}.  };
%%%%
\begin{align}
	\sigma\equiv(-1)^{F_L}\otimes({\bf -1}_R)^{\otimes4},
\label{asym-o}
\end{align}
%%%%
where $(-1)^{F_L}$ is space-time fermion number on the left-mover and $({\bf -1}_R)^{\otimes4}$ denotes $X^i_R\rightarrow-X^i_R$, $\psi^i_R\rightarrow-\psi^i_R\;(i=6,\dots,9)$. 
%%%%
The total $\bz_2\times \bz_2$-oribifold 
is simply identified as 
$$
\left[ T^4/\cI_4 \right]/\sigma \cong \mbox{K3}/\sigma .
$$

%%%%%%%%%%%%%%%%%%%%%%%%%%%%%%%%%%%%%%%%%%%%%%%%%%%%%%%%%%%%%%%%%%

~

\subsection{The torus partition function}

Let us first consider the partition function of torus $T^4\equiv T^4[SO(8)]$:
%%%%
\begin{align}
	Z^{T^4}(\tau,\bar{\tau}) = \frac{1}{2} \left\{ \left|\frac{\theta_3}{\eta}\right|^8 + \left|\frac{\theta_4}{\eta}\right|^8 + \left|\frac{\theta_2}{\eta}\right|^8 \right\},
\label{bosonT4}
\end{align}
and we introduce the symbol
\begin{align}
\cJ(\tau) := \frac{1}{\eta^4} \left[\th_3^4 - \th_4^4 -\th_2^4\right](\tau) (\equiv 0). 
%	\mathcal{J}(\tau) := \frac{1}{\eta(\tau)^4} \left\{ \theta_3(\tau)^4 - \theta_4(\tau)^4 - \theta_2(\tau)^4 \right\} (\equiv0),
%\frac{1}{2\eta(\tau)^4} \left\{ \theta_3(\tau)^4 - \theta_4(\tau)^4 - \theta_2(\tau)^4 \right\} (\equiv0),
\label{fermion}
\end{align}
as the supersymmetric chiral blocks for free fermion. Therefore, the modular invariant partition function on $T^4$ compactification is written as
\begin{align}
	Z(\tau,\bar{\tau}) = Z^{6d}_{\text{bosonic}}(\tau,\bar{\tau})\, \frac{1}{4} Z^{T^4}(\tau,\bar{\tau}) \mathcal{J}(\tau) \overline{\mathcal{J}(\tau)},
\label{torus}
\end{align}
where $Z^{6d}_{\text{bosonic}}(\tau,\bar{\tau})$ denotes the partition function of the bosonic sector of uncompactified space-time $\bsr^{5,1}$. 
%$\mathcal{J}(\tau)$  and $\overline{\cJ(\tau)}$ are left- and right-moving fermionic blocks.

%%%%%%%%%%%%%%%%%%%%%%%%%%%%%%%%%%%%%%%%%%%%%%%%%%%%%%%%%%%%%%%%%%%%%%%%%
%%%%%%%%%%%%%%%%%%%%%%%%%%%%%%%%%%%%%%%%%%%%%%%%%%%%%%%%%%%%%%%%%%%%%%%%%
%%%%%%%%%%%%%%%%%%%%%%%%%%%%%%%%%%%%%%%%%%%%%%%%%%%%%%%%%%%%%%%%%%%%%%%%%

Now, the total modular invariant of the $\bz_2 \times \bz_2 \equiv \left\langle \cI_4,  \sigma \right\rangle $-orbifold of $T^4$ is written as 
%%%
\begin{align}
Z(\tau,\bar{\tau}) & = Z^{6d}_{\text{bosonic}}(\tau,\bar{\tau}) \,
 \frac{1}{16} 
%\sum_{s_L,s_R}\, 
\sum_{\al,\beta, a,b} \, 
%\delta(s_L) \delta(s_R) \,
%Z^{(s_L,s_R)}_{(\al,\beta), (a,b)} (\tau, \bar{\tau}).
Z_{(\al,\beta), (a,b)} (\tau, \bar{\tau}).
\label{pf general}
\end{align}
Here, $(\al,\beta) , (a,b) \in \bz_2\times \bz_2$ characterize the $\cI_4$- and $\sigma$-twists.
The building block $Z_{(\al,\beta), (a,b)} (\tau, \bar{\tau})$ satisfies the modular covariance relation;
\begin{align}
& Z_{(\al,\beta), (a,b)} (\tau+1, \bar{\tau}+1) = Z_{(\al,\al+\beta), (a,a+b)} (\tau, \bar{\tau}), 
\nn
&  Z_{(\al,\beta), (a,b)} \left(-\frac{1}{\tau}, -\frac{1}{\bar{\tau}}\right) =  Z_{(\beta,\al), (b,a)} (\tau, \bar{\tau}).
\label{mod cov general}
\end{align}
The overall factor $1/16$ has been determined from the order of  orbifold group as well as the chiral GSO projection\footnote
 {Here, $4 \equiv 2 \cdot 2$ originates from the chiral GSO-projection, 
while  each of the twist labels $(\al,\beta)$, $(a,b)$ has periodicity 2 as explicitly confirmed.
This fact is not necessarily self-evident because of the {\em asymmetry\/} of orbifold action of our interests.}. 

%%%%%%%%%%%%%%%%%%%%%%%%%%%%%%%%

It is convenient to decompose the total partition function \eqn{pf general} into the contributions from the several `modular orbits', each of which is separately modular invariant;
\begin{align}
& Z(\tau,\bar{\tau})  = Z^{6d}_{\text{bosonic}}(\tau,\bar{\tau}) \frac{1}{16} \left\{ Z[{\bf 1}] + Z[\cI_4]+ Z[\sigma]  + Z[\sigma \cI_4] + Z[\sigma, \cI_4] \right\},
\label{orbit decomp}
\end{align}  
where we define the each modular orbit as 
%%%%%%%%%%%%%%%%%%%%%
\begin{align}
&  Z[{\bf 1}] \equiv Z_{(0,0),(0,0)}, \hspace{1cm} Z[\cI_4] \equiv \sum_{(\al,\beta) \neq (0,0)} \, Z_{(\al,\beta),(0,0)}, \hspace{1cm} 
Z[\sigma] \equiv \sum_{(a,b) \neq (0,0)} \, Z_{(0,0), (a,b)}, 
\nn
& Z[\sigma \cI_4] \equiv \sum_{(a,b) \neq (0,0)} \, Z_{(a,b),(a,b)},
\hspace{1cm} 
Z[\sigma, \cI_4] \equiv \sum_{\stackrel{(a,b) \neq (\al,\beta)}{(\al,\beta) \neq (0,0), (a,b) \neq (0,0)}} \, Z_{(\al,\beta),(a,b)}.
\label{def orbits}
\end{align}
The untwisted sector corresponds to 
\begin{align}
Z[{\bf 1}] =  Z^{T^4}(\tau,\bar{\tau}) \mathcal{J}(\tau) \overline{\mathcal{J}(\tau)},
\label{Z1}
\end{align}
and 
$Z[\cI_4]$, $Z[\sigma]$ and $Z[\sigma \cI_4]$ are identified as  the twisted sectors for the involutions $\cI_4$, $\sigma$, $\sigma \cI_4$.
The explicit forms of these functions are summarized in appendix \ref{app:orbit}.
For instance, 
the compactification onto the asymmetric orbifold $T^4/\sigma$ is described by the modular invariant 
$$
Z^{T^4/\sigma}(\tau,\bar{\tau}) \equiv Z^{6d}_{\text{bosonic}}(\tau,\bar{\tau}) \cdot \frac{1}{8} \left[Z[{\bf 1}] + Z[\sigma]\right] ,
$$
while the K3 ($\cong T^4/\cI_4$)-compactification is expressed as 
$$
Z^{\msc{K3}}(\tau,\bar{\tau}) \equiv Z^{6d}_{\text{bosonic}}(\tau,\bar{\tau}) \cdot \frac{1}{8} \left[Z[{\bf 1}] + Z[\cI_4]\right] .
$$
%%%%%

The remaining orbit $Z[\sigma, \cI_4]$ plays a crucial role in this work, and it is evaluated as 
\begin{align}
Z[\sigma, \cI_4] & = \frac{1}{2} \left(\frac{2\eta}{\th_4}\right)^2 \left(\frac{\th_3^2 \th_2^2}{\eta^4} + \frac{\th_2^2 \th_3^2}{\eta^4}\right)
\cdot \overline{
\left(\frac{2\eta}{\th_3}\right)^2 \left(\frac{\th_4^2 \th_2^2}{\eta^4} - \frac{\th_2^2 \th_4^2}{\eta^4}\right)
} + \cdots
\nn
& \equiv \frac{1}{2} \frac{1}{|\eta|^{16}} \left(\th_3^4 \th_2^4 + \th_2^4 \th_3^4 \right) \cdot \overline{\left(\th_4^4 \th_2^4 - \th_2^4 \th_4^4 \right)} + \cdots,
\label{orbit sigma cI4 0}
\end{align}
where we explicitly exhibit the the contribution  
%twisted sector ' $\dsp  {\scriptstyle {\sigma}} \underset{\cI_4}{\square} $ ', that is, 
$ \tr_{\cI_4\msc{-twsited}} \left[\sigma q^{L_0-\frac{c}{24}}\overline{q^{\tL_0-\frac{c}{24}}} \right]$,
and the omitted terms $\cdots $ are uniquely determined by modular transformations. 
%%%%%%%%%%%%%
To derive the second line, we made use of the familiar identity $\th_2 \th_3 \th_4 = 2\eta^3$.
%%%%%%%%%%%%%
We will later discuss various modifications of this orbit by including the discrete torsion.

It is worthwhile to note the massless spectrum in the $\cI_4$-twisted sector, which is directly extracted from the combination of orbits
$Z[\cI_4] + Z[\sigma, \cI_4]$.
After making straightforward computations, we can evaluate the relevant contributions as follows;
%%%%%%%%%%
%%%%%%%%%%
%%%%%%%%%%
\begin{align}
& 
%\left[ Z[\cI_4] + Z[\sigma, \cI_4] \right]_{\cI_4\msc{-twisted}, \msc{NS-NS}}  
\left[ \bigbox{{\bf 1}}{\cI_4} + \bigbox{\cI_4}{\cI_4}
+ \bigbox{\sigma}{\cI_4} + \bigbox{\sigma\cI_4}{\cI_4} \right]_{\msc{NS-NS}}
= \frac{1}{|\eta|^{16}}
\left|\th_2\right|^8\left\{ \left|\th_3\right|^8 + \left|\th_4\right|^8\right\} + \frac{1}{2} \frac{1}{|\eta|^{16}}
\left|\th_2\right|^8\left\{ \th_3^4 \overline{\th_4^4} + \th_4^4 \overline{\th_3^4} \right\},
\nn
& 
%\left[ Z[\cI_4] + Z[\sigma, \cI_4] \right]_{\cI_4\msc{-twisted}, \msc{R-R}}  = \frac{1}{|\eta|^{16}}
\left[ \bigbox{{\bf 1}}{\cI_4} + \bigbox{\cI_4}{\cI_4}
+ \bigbox{\sigma}{\cI_4} + \bigbox{\sigma\cI_4}{\cI_4}  \right]_{\msc{R-R}}
= \frac{1}{|\eta|^{16}}\left|\th_2\right|^8\left\{ \left|\th_3\right|^8 + \left|\th_4\right|^8\right\} - \frac{1}{2} \frac{1}{|\eta|^{16}}
\left|\th_2\right|^8\left\{ \th_3^4 \overline{\th_4^4} + \th_4^4 \overline{\th_3^4} \right\}.
\label{Ztorus 0}
\end{align}
%%%%%%%%%
%%%%%%%%%
%%%%%%%%%
Here we adopted a schematic (probably, standard) expression to exhibit the traces with various twisted sectors. 
For instance, 
$$
\left[ \bigbox{\sigma}{\cI_4} \right]_{\msc{NS-NS}} \equiv \tr_{\cI_4\msc{-twisted sector}, \, \msc{NS-NS}} 
\left[\sigma\, q^{L_0-\frac{c}{24}} \overline{q^{\tL_0-\frac{c}{24}}}\right].
$$
%%%%%%%%%
The numbers of NS-NS and R-R massless states in the $\cI_4$-twisted sector are read from the $q$-expansions of  \eqn{Ztorus 0}, which 
we denote as `$N_{\cI_4\msc{-tw}}^{(\sNS,\sNS)}$' and `$N_{\cI_4\msc{-tw}}^{(\sR,\sR)}$' from now on.
Needless to say, no tachyonic states exist in this case, and the leading terms of $q$-expansions correspond to the massless states. 
We so obtain
\begin{align}
N_{\cI_4\msc{-tw}}^{(\sNS,\sNS)} =48, \hspace{1cm} N_{\cI_4\msc{-tw}}^{(\sR,\sR)} =16,
\label{case 1 massless B}
\end{align}
On the other hand, since having the unbroken SUSY in the right-mover, 
the numbers of massless fermions in  the $\cI_4$-twisted sector are just found to be
\begin{align}
N_{\cI_4\msc{-tw}}^{(\sNS,\sR)} =48, \hspace{1cm} N_{\cI_4\msc{-tw}}^{(\sR,\sNS)} =16.
\label{case 1 massless F}
\end{align}
%%%%

~

%%%%%%%%%%%%%%%%%%%%%%%%%%%%%%%%%%%%%%%%%%%%%%%%%%%%%%%%%%%%%%%%%%
%%%%%%%%%%%%%%%%%%%%%%%%%%%%%%%%%%%%%%%%%%%%%%%%%%%%%%%%%%%%%%%%%%

\subsection{The cylinder partition function}

In this subsection, we analyze the cylinder partition functions of which both ends are attached to the same D-brane by using the method of boundary states. 
First, we introduce the `bulk-type branes' defined in $\mbox{K3} \cong T^4 / \cI_4$
%\footnote{We call the boundary state without the twisted sector the bulk-type brane in contrast with the fractional brane.} 
as the (BPS) GSO-projected boundary state $\dket{B}_{\msc{bulk}}$ composed only of the $\cI_4$-untwisted sector;
\begin{align}
\dket{B}_{\msc{bulk}}= & \cN \cP_{\msc{GSO}} \left[
\exp\left(\sum_{n\in \bsz}-\frac{\ep_{(\mu)}}{n}\alpha^\mu_{-n}\tilde{\alpha}_{\mu, -n}
-i \sum_{r\in \bsz+\frac{1}{2}} \ep_{(\mu)} \psi^{\mu}_{-r}\tilde{\psi}_{\mu,-r}\right)\dket{B_0}_{\sNS\sNS}
\right.
\nn
& \hspace{2cm} \left. + \exp\left(\sum_{n\in \bsz}-\frac{\ep_{(\mu)}}{n}\alpha^\mu_{-n}\tilde{\alpha}_{\mu, -n}
-i \sum_{r\in \bsz}\ep_{(\mu)} \psi^{\mu}_{-r}\tilde{\psi}_{\mu,-r}\right)
\dket{B_0}_{\sR\sR}\right],
%	\dket{B}_{\msc{bulk}}=\cN \cP_{\msc{GSO}}
%\exp\left(\sum_{n\in \bsz}-\frac{1}{n}\alpha^\mu_{-n}\tilde{\alpha}_{\mu, -n}-i\eta\sum_{r\in \bsz+\nu}\psi^{\mu}_{-r}\tilde{\psi}_{\mu,-r}\right)\dket{B_0},
\label{BPSbs}
\end{align}
%%%%
where $\alpha^\mu_n$, $\tilde{\alpha}^\mu_n$ are the left- and right-moving modes of $X^\mu$, and $\psi^{\mu}_r$ and $\tilde{\psi}^\mu_r$ are the modes of $\psi^{\mu}$ and $\tilde{\psi}^\mu$, respectively. 
%$\mu$ denotes the direction of space-time dimension. 
The sign factor $\ep_{(\mu)} = \pm 1$ depends on whether the $\mu$-direction is Dirichlet or Neumann, which won't play any roles  of importance in the following arguments.   
%%%%
%%%%
%$\dket{\bf{0}}$ 
$\dket{B_0}$
denotes the zero-mode component, which should be $\cI_4$-invariant, and  $\cN$ is the normalization factor to be determined by the open-closed correspondence. 
%%%
$\cP_{\msc{GSO}}$ denotes the standard  GSO projection operator which chirally acts on the, say, left-mover only.
%%%
In other words, \eqn{BPSbs} is nothing but the $\cI_4$-invariant combinations of the BPS boundary states under the $T^4$-compactification.

%%%%%%%%%%%%%%%%%%%%%%%%%%%%%%%%%%%%%%%%%%

The relevant cylinder partition function, which we call  
the `self-overlap' of boundary states is just evaluated  as 
\begin{align}
	{}_{\msc{bulk}}\dbra{B} e^{-\pi s H^{(c)}} \dket{B}_{\msc{bulk}} = 0,
\label{BPSol}
\end{align}
where $s$ and $H^{(c)}\equiv L_0+\tL_0- \frac{c}{12}$ denotes the closed string modulus and Hamiltonian for the cylinder. 
Then we obtain the vanishing self-overlap 
because $\dket{B}_{\msc{bulk}}$ is BPS.

%%%%%%%%%%%%%%%%%%%%%%%%%%%%%%%%%%%%%%%%%%%%%%%%%%%%%%%%%%%%%

In the previous paper \cite{SSU}, 
%we discussed the self-overlap in the asymmetric orbifolds by the involution $\sigma$ for the D-branes  described by the 
we studied the D-branes in the asymmetric orbifold by the involution $\sigma$
described by the boundary states of the form;
\begin{equation}
\dket{\cB} \equiv  \sqrt{2} \cP_{\sigma} \dket{B},
\label{cB}
\end{equation}
where $\dket{B}$ denotes an arbitrary BPS brane in the unorbifolded theory and $\dsp \cP_{\sigma} \equiv \frac{1+\sigma}{2}$ is the $\sigma$-projection. 
The numerical factor $\sqrt{2}$ is necessary due to the Cardy condition. 
As already mentioned, such boundary states $\dket{\cB}$ never satisfy the BPS equation, since we  at most possess the chiral SUSY after making the $\sigma$-orbifolding.
Nonetheless, as addressed in \cite{SSU}, the self-overlaps vanishes rather generically
%%%%%
\begin{align}
\dbra{\cB} e^{-\pi s H^{(c)}} \dket{\cB} = 
	\dbra{B} e^{-\pi s H^{(c)}} \dket{B} + \dbra{B} \sigma e^{-\pi s H^{(c)}} \dket{B}= 0.
\label{Bulktype}
\end{align}
%%%%
To be more precise, it is true for any boundary states written as \eqn{cB} with the `bulk-type' branes \eqn{BPSbs} in $T^4/\cI_4$. 

%%%%%%%%%%%%%%%%%%%%%%%%%%%%%%%%%%%%%%%%%%%%%%%%%%%%%%%%%%%%%%%%%%%%%%%%
%%%%%%%%%%%%%%%%%%%%%%%%%%%%%%%%%%%%%%%%%%%%%%%%%%%%%%%%%%%%%%%%%%%%%%%%

However, this is not the whole story. One can still accommodate the fractional branes in the orbifold  $T^4/\cI_4$,
that is, the boundary states including the contributions from the $\cI_4$-twisted sector\footnote
   {It is obvious that the $\sigma$- and $\sigma \cI_4$-twisted sectors cannot contribute to boundary states due to their asymmetric actions. }. 
%%%%
To do so, we have to consider the boundary states of the type \eqn{cB} 
with $\dket{B} \equiv \dket{B}_{\text{frac}}$ given by (for the transverse part);
\begin{align}
\begin{aligned}
\dket{B}_{\text{frac}} & = \dket{B}_{U} + \dket{B}_{T}
%\\ &
\equiv \dket{B}_{\sNS\sNS;U}+\dket{B}_{\sR\sR;U}+\dket{B}_{\sNS\sNS;T}+\dket{B}_{\sR\sR;T}, \\
& \dket{B}_{\sNS\sNS;U} = \cN  \cP_{\msc{GSO}} \exp \left(\sum_{n\in \bsz} -\frac{ \ep_{(\mu)}}{n}\alpha^\mu_{-n}\tilde{\alpha}_{\mu, -n} 
-i\sum_{r\in \bsz+\frac{1}{2}} \ep_{(\mu)}\psi^{\mu}_{-r}\tilde{\psi}_{\mu,-r}\right)\dket{B_0}_{\sNS\sNS;U}, \\
&\dket{B}_{\sR\sR;U} = \cN \cP_{\msc{GSO}} 
\exp\left(\sum_{n\in \bsz} -\frac{\ep_{(\mu)}}{n}\alpha^\mu_{-n}\tilde{\alpha}_{\mu, -n} -i\sum_{r\in \bsz} \ep_{(\mu)} \psi^{\mu}_{-r}\tilde{\psi}_{\mu,-r}\right)
\dket{B_0}_{\sR\sR;U}, \\
&\dket{B}_{\sNS\sNS;T} = 
% \tilde{\cN} 
\cN \cP_{\msc{GSO}} \exp\left(\sum_{n\in \bsz+a} -\frac{\ep_{(\mu)}}{n}\alpha^\mu_{-n}\tilde{\alpha}_{\mu, -n}
-i\sum_{r\in \bsz+b}\ep_{(\mu)} \psi^{\mu}_{-r}\tilde{\psi}_{\mu,-r}\right)\dket{B_0}_{\sNS\sNS;T}, \\
&\dket{B}_{\sR\sR;T} = 
%% \tilde{\cN}
\cN \cP_{\msc{GSO}}
\exp\left(\sum_{n\in \bsz+a} -\frac{\ep_{(\mu)}}{n}\alpha^\mu_{-n}\tilde{\alpha}_{\mu, -n} -i\sum_{r\in \bsz+a} \ep_{(\mu)} \psi^{\mu}_{-r}\tilde{\psi}_{\mu,-r}\right)
\dket{B_0}_{\sR\sR;T},
\label{frac}
\end{aligned}
\end{align}
with
\begin{align}
	a=\left\{
		\begin{array}{c}
			0\;\;\;:X^{2,\dots,5} \\
			\frac{1}{2}\;\;\;:X^{6,\dots,9} \\
		\end{array} 
	\right.\;\;\;\;\;b=\left\{
		\begin{array}{c}
			\frac{1}{2}\;\;\;:X^{2,\dots,5} \\
			0\;\;\;:X^{6,\dots,9} \\
		\end{array} 
	\right.,
\label{integer}
\end{align}
where $U$ and $T$ denotes the untwisted sector and the twisted sector, and the untwisted sectors have the momentum along the $i$th compact direction. 
%%%%%%%%%%%%%%%%%

Now, the self-overlap of boundary state $\dket{\cB} \equiv \sqrt{2} \cP_{\sigma} \dket{B}_{\msc{frac}}$ is evaluated as follows;
\begin{equation}
\dbra{\cB} e^{-\pi s H^{(c)}} \dket{\cB} = {}_{\msc{frac}} \dbra{B} e^{-\pi s H^{(c)}} \dket{B}_{\msc{frac}} + {}_{\msc{frac}} \dbra{B} \sigma e^{-\pi s H^{(c)}} \dket{B}_{\msc{frac}},
\label{Zcyl frac 0}
\end{equation}
%%%%
and 
\begin{align}
\begin{aligned}
	{}_{\msc{frac}}\dbra{B} e^{-\pi s H^{(c)}} \dket{B}_{\text{frac}} &=  \frac{1}{2}
Z_{\msc{zero-modes}} (is) \cdot 
\frac{1}{\eta(is)^8} \cdot \frac{1}{\eta(is)^4} \left[\th^4_3 - \th_4^4- \th_2^4\right](is) \\
%\left[ \left(\frac{\theta_3(is)}{\eta(is)}\right)^4 - \left(\frac{\theta_4(is)}{\eta(is)}\right)^4 - \left(\frac{\theta_2(is)}{\eta(is)}\right)^4 \right]\\
	&\hspace{10pt} +
% \tilde{\cN}^2 
\frac{1}{2} \frac{1}{\eta(is)^4} \cdot \left(\frac{2 \eta}{\th_4}\right)^2(is) \cdot \frac{1}{\eta(is)^4} \left[\th_3^2 \th_2^2 - \th_2^2 \th_3^2\right] (is) = 0.
%\left[ \frac{\theta_3(is)^2\theta_2(is)^2}{\eta(is)^4} 
%- \frac{\theta_2(is)^2\theta_3(is)^2}{\eta(is)^4} \right] \equiv 0,
%\\
%	&\sim {\scriptstyle (\bold{1},\bold{1})} \underset{(\bold{1},\bold{1})}{\square} + \;{\scriptstyle {(\bold{1},\bold{1})}} \underset{(\cI_4,\bold{1})}{\square} = 0,
\label{Z-K3frac0}
\end{aligned}
\end{align}
because of the BPS property of $\dket{B}_{\text{frac}}$ in the vacuum $\mbox{K3} \cong T^4/\cI_4$. 
%%%%%%%%%%%%%%%%%%%%%%%%%%%%%%%%%%%
%%%%%%%%%%%%%%%%%%%%%%%%%%%%%%%%%%%
(The factor $1/\eta(is)^4$ comes from the bosonic transverse part of $\br^{5,1}$-sector.)
%%%%%%%%%%%%%%%%%%%%%%%%%%%%%%%%%%%
%%%%%%%%%%%%%%%%%%%%%%%%%%%%%%%%%%%
On the other hand, taking account of $\sigma \equiv (-1)^{F_L} \otimes (\-_R)^{\otimes 4}$, 
we obtain 
\begin{align}
\begin{aligned}
	{}_{\msc{frac}}\dbra{B} \sigma\, e^{-\pi s H^{(c)}} \dket{B}_{\text{frac}} &= 
%\cN^2 
\frac{1}{2} \frac{1}{\eta(is)^4} \cdot \left(\frac{2 \eta}{\th_2}\right)^2(is)  \cdot \frac{1}{\eta(is)^4} \left[ \th_3^2 \th_4^2 - \th_4^2 \th_3^2\right](is) \\
%\left[ \frac{\theta_3(is)^2\theta_4(is)^2}{\eta(is)^4} - \frac{\theta_4(is)^2\theta_3(is)^2}{\eta(is)^4} 
%+ \frac{\theta_2(is)^2\theta_1(is)^2}{\eta(is)^4} \right]\\
	&\hspace{10pt} + 
%\tilde{\cN}^2 
\frac{1}{2}
\frac{1}{\eta(is)^4} \cdot \left(\frac{2 \eta}{\th_3}\right)^2(is) \cdot \frac{1}{\eta(is)^4} \left[ \th_4^2 \th_2^2 + \th_2^2 \th_4^2\right](is)
%\frac{\theta_4(is)^2\theta_2(is)^2}{\eta(is)^4} + \frac{\theta_2(is)^2\theta_4(is)^2}{\eta(is)^4} \right] 
\neq 0.
%\\
%	&\sim {\scriptstyle (\bold{1},\sigma)} \underset{(\bold{1},\bold{1})}{\square} + \;{\scriptstyle {(\bold{1},\sigma)}} \underset{(\cI_4,\bold{1})}{\square} \neq 0.
\label{Z-K3frac1}
\end{aligned}
\end{align}
%%%%%
In this way, we have shown that the self-overlap \eqn{Zcyl frac 0} does not vanish for generic fractional branes. 
%%%%
%%%%
We will discuss the possibility to change this feature by extending the $\sigma$-orbifold so as to include the discrete torsion in the next section. 
%%%%
%%%%

It is also valuable to examine the massless open string spectrum associated to such non-BPS branes. 
Indeed, by combining \eqn{Z-K3frac0} and \eqn{Z-K3frac1} and making the modular transformation $s = 1/t $, 
we obtain 
\begin{align}
Z^{\msc{cyl}}_{\msc{frac}} (i/t) & \equiv {}_{\msc{frac}} \dbra{B} e^{-\pi s H^{(c)}} \dket{B}_{\msc{frac}} + {}_{\msc{frac}} \dbra{B} \sigma e^{-\pi s H^{(c)}} \dket{B}_{\msc{frac}}
\nn
%& =
%\frac{1}{2} \frac{1}{\eta(is)^{12}} \cdot \left[ Z_{\msc{zero-modes}}(is)\cdot 
%\left(\th_3^4 - \th_4^4 - \th_2^4\right) (is)
%+ \left(\th_3^4 \th_4^4- \th_4^4 \th_3^4\right)(is)
%\right.
%\nn
%& \hspace{2cm}
%\left. + \left(\th_3^4 \th_2^4- \th_2^4 \th_3^4\right)(is)  +  \left(\th_4^4 \th_2^4+ \th_2^4 \th_4^4\right) (is) \right] 
%\nn
& =
\frac{1}{2} \frac{1}{t^2 \eta(it)^{12}} \cdot \left[ \widetilde{Z}_{\msc{zero-modes}}(it)\cdot 
\left(\th_3^4 - \th_4^4 - \th_2^4\right) (it)
+ \left(\th_3^4 \th_4^4- \th_4^4 \th_3^4\right)(it)
\right.
\nn
& \hspace{2cm}
%\left. + \left(\th_3^4 \th_2^4- \th_2^4 \th_3^4\right)(it)  +  \left(\th_4^4 \th_2^4+ \th_2^4 \th_4^4\right) (it) \right] 
\left. + \th_2^4 \left(\th_3^4 + \th_4^4\right) (it) - \th_2^4 \left(\th_3^4 - \th_4^4 \right)(it)   \right] ,
\label{eval Zcyl 0}
\end{align} 
where we simply wrote $Z_{\msc{zero-modes}}(i s) = t^2 \widetilde{Z}_{\msc{zero-modes}}(it)$.
The 3rd and 4th terms appearing in the second line of \eqn{eval Zcyl 0} are identified with the twisted open strings in the NS and R-sectors respectively. 
Therefore, when comparing with the bulk-type branes, the numbers of extra massless states in the NS and R open string sectors are evaluated as 
\begin{equation}
n^{(\sNS)}_{\msc{tw}} = 16, \hspace{1cm} n^{(\sR)}_{\msc{tw}} =0.  
\end{equation}

%%%%%%%%%%%%%%%%%%%%%%%%%%%%%%%%%%%%%%%%%%%%%%%%%%%%%%%%
%%%%%%%%%%%%%%%%%%%%%%%%%%%%%%%%%%%%%%%%%%%%%%%%%%%%%%%%
%%%%%%%%%%%%%%%%%%%%%%%%%%%%%%%%%%%%%%%%%%%%%%%%%%%%%%%%
%%%%%%%%%%%%%%%%%%%%%%%%%%%%%%%%%%%%%%%%%%%%%%%%%%%%%%%%

\section{Asymmetric orbifolds with discrete torsions}
\label{sec dtorsion}

\subsection{Inclusion of discrete torsion}

In the previous section, we have studied the asymmetric orbifold  written schematically as 
\begin{align}
	\left(T^4/\cI_4\right)/\sigma.
\label{non-discback}
\end{align}
%%%
We would like to discuss various extensions of this model so as to include the 
relative phase factors among the relevant orbifold twists, $\cI_4$ and $\sigma$, 
which are broadly known as the 
`discrete torsion' \cite{Vafa:1986wx,Vafa:1994rv,Gaberdiel:2000fe}.
%%%%%%%%%%%
We  especially investigate whether we gain the vanishing self-overlaps associated to the fractional branes in such modified vacua.
%%%%%%%%%%%
It is a novel point here that we shall adopt the discrete torsion that {\em depend on the spin structures}.
In other words we allow relative phases among the orbifold twists by $\cI_4$ and $\sigma$
{\em as well as the GSO twisting}, which can be also regarded as a $\bz_2$-orbifolding. 

%%%%%%%%%%%%%%%%%%%%%%%%%%%%%%%%%%%%%%%%%%%%%%%%%%%%%%%%%%

Thus, the most general form of torus partition function is written schematically as 
\begin{align}
Z(\tau,\bar{\tau}) & = Z^{6d}_{\text{bosonic}}(\tau,\bar{\tau}) \frac{1}{16} \sum_{s_L,s_R}\, \sum_{\al,\beta, a,b} \, \delta(s_L) \delta(s_R) \,
\ep^{(s_L,s_R)}_{(\al,\beta), (a,b)}\,
Z^{(s_L,s_R)}_{(\al,\beta), (a,b)} (\tau, \bar{\tau}).
\label{pf dt general}
\end{align}
In this expression, $(\al,\beta)$ and $(a,b)$ again characterize the $\cI_4$ and $\sigma$ twists, while $s_L,s_R \equiv \NS, \tNS, \R$ express the chiral spin structures. 
Both of the total building block $Z^{(s_L,s_R)}_{(\al,\beta), (a,b)} (\tau, \bar{\tau})$ and the phase factor (discrete torsion) $\ep^{(s_L,s_R)}_{(\al,\beta), (a,b)}$ 
should keep the modular covariance. Especially we require $\ep^{(s_L,s_R)}_{(\al,\beta), (a,b)}$ to be invariant on each modular orbit.
$\delta(s_L)$, $\delta(s_R)$ are the standard sign factors associated to the chiral GSO projection;
$\delta(\NS) =1$, ~ $\delta(\tNS) = \delta(\R)=-1$. (We omit the factor $\delta(\tR) = \pm 1$ due to the existence of $\br^{5,1}$-sector.)
%The overall factor $1/16$ has been determined from the periodicity
%\footnote
% {Here, $4 \equiv 2 \cdot 2$ originates from the chiral GSO-projection, 
%while  each of the twist labels $(\al,\beta)$, $(a,b)$ has periodicity 2 as explicitly confirmed.
%This fact is not necessarily self-evident because of the {\em asymmetry\/} of orbifold action of our interests.} of the (generic) building blocks 
%$Z^{(s_L,s_R)}_{(\al,\beta), (a,b)} (\tau, \bar{\tau})$.

%%%%%%%%%%%%%%%%%%%%%%%%%%%%%%%%%%%%%%%%%%

To avoid unessential complexities, 
we further assume that the modular orbits $Z[{\bf 1}]$, $Z[\cI_4]$, and $Z[\sigma]$ remain unchanged. 
This assumption obviously implies $\ep^{(s_L,s_R)}_{(0,0), (a,b)} =1$, $(\any a,b)$.
%%%%%%%%%%%%%%%%%%%%%%%%%%%%%%%%%%%%%%%%%%%%%%%%%%%%%%%%%%%%%%%%%%%%%%%%%%%%%%%%%%%%%%%%%%%%%%%%%%%%%%%%%%%
%\begin{itemize}
%%\item Before making the $\sigma$-orbifolding,  the $\cI_4$-orbifolding describes 
%%the supersymmetric K3-compactification, that is,  the type II string vacua with $\cN=(1,1)$ SUSY in 6-dim. space-time. 
%%The simplest solution of this requirement is of course 
%%$\ep^{(s_L,s_R)}_{(\al,\beta), (0,0)} =1$. 
%%
%\item 
%$\sigma $ acts in the same way as before on the $\cI_4$-untwisted sector, leaving 
%$Z[{\bf 1}]$, $Z[\sigma]$ unchanged. 
%This just means that 
%$
%\ep^{(s_L,s_R)}_{(0,0), (a,b)} =1.
%$
%
%\item In the $\sigma$-untwisted sector, which means $(a,b)=(0,0)$, the $\cI_4$-orbifolding always describes 
%the supersymmetric K3-compactification, that is,  the type II string vacua with $\cN=(1,1)$ SUSY in 6-dim. space-time. 
%%Note that we still have the possibilities;
%%$\ep^{(s_L,s_R)}_{(0,1), (0,0)} =1$, 
%%$\ep^{(s_L,s_R)}_{(0,1), (0,0)} =(-1)^{F_L}$,
%%$\ep^{(s_L,s_R)}_{(0,1), (0,0)} =(-1)^{F_R}$,
%%$\ep^{(s_L,s_R)}_{(0,1), (0,0)} =(-1)^{F_L+F_R}$,
%
%\end{itemize}
%%%%%%%%%%%%%%%%%%%%%%%%%%%%%%%%%%%%%%%%%%%%%%%%%%%%%%%%
However, 
we still have the possibilities to replace $\cI_4$ with $(-1)^{F_L} \cI_4$, $(-1)^{F_R} \cI_4$, or $(-1)^{F_L+F_R} \cI_4$,
all of which lead to an equal modular invariant $Z[\cI_4]$ because 
$$
Z^{(\sR,*)}_{(0,1), (0,0)} (\tau, \bar{\tau}) \equiv Z^{(*,\sR)}_{(0,1), (0,0)} (\tau, \bar{\tau}) \equiv 0.
$$
This `degeneracy'  is partly removed if considering the sectors $(a,b)=(\al,\beta)$, in other words, if focusing on the modular orbit $Z[\sigma \cI_4]$.
We  obtain two inequivalent vacua by choosing the phase factor as 
{\bf (1)} $\ep^{(s_L,s_R)}_{(0,1),(0,1)} = 1$, or 
{\bf (2)} $\ep^{(s_L,s_R)}_{(0,1),(0,1)} = (-1)^{F_R}$.
(The other phase factors $\ep^{(s_L,s_R)}_{(a,b), (a,b)}$ are uniquely determined by the modular invariance.)
%%%%
The first case corresponds to the $\cI_4$ (or $(-1)^{F_L} \cI_4$)-orbifolding, whereas the second one does to $\tcI_4 \equiv (-1)^{F_R} \cI_4$ (or $(-1)^{F_L+F_R} \cI_4$).
We note that $\sigma \cI_4 \equiv  (-1)^{F_L} (\-_L)^{\otimes 4}$ and $\sigma \tcI_4 \equiv (-1)^{F_L+F_R} (\-_L)^{\otimes 4}$ yield different modular orbits;
$$
Z[\sigma \cI_4] \neq Z[\sigma \tcI_4].
$$ 
In fact, the original orbit $Z[\sigma \cI_4]$ respects the right-moving chiral SUSY, whereas the modified one $Z[\sigma \tcI_4]$ breaks the right-moving SUSY 
although the  bose-fermi cancellation happens in the left-mover. 
See appendix \ref{app:orbit} for the explicit forms of these orbits.

For each of $\cI_4$ and $\tcI_4$-orbifolding, we can still include the non-trivial phases $\ep(s_L,s_R)\equiv \ep^{(s_L,s_R)}_{(1,0), (0,1)}$.
This uniquely determines the phase factors for the sectors with $(\al,\beta) \neq (0,0)$, $(a,b)\neq (0,0)$, and $(\al,\beta) \neq (a,b)$, which lie on the same modular orbit.
We first count how many choices are allowed for this phase. 
%%%%
The conformal invariance implies that the it should be factorized as $\ep(s_L,s_R) \equiv \ep_L(s_L)\ep_R(s_R)$, and 
$\ep_L (s_L)= \pm 1$, $\ep_R (s_R) = \pm 1$ can be independently chosen. 
However, 
since
$$
Z^{(\stNS,*)}_{(1,0), (0,1)} (\tau, \bar{\tau}) \equiv 0, 
\hspace{1cm}
Z^{(*,\sNS)}_{(0,1), (0,0)} (\tau, \bar{\tau}) \equiv 0,
$$
holds, the relevant phases are determined by choosing $\ep_L(\NS)$, $\ep_L(R)$, $\ep_R(\tNS)$ and $\ep_R(\R)$. 
We also note a manifest symmetry for $\ep(s_L,s_R)$ 
by $\left(\ep_L(s_L), \ep_R(s_R)\right) \, \rightarrow \, \left(-\ep_L(s_L), - \ep_R(s_R)\right)$.
Thus, we end up with the $\dsp 2^4 \times \frac{1}{2} =8$ independent choices of $\ep(s_L,s_R)$, explicitly summarized in the table \ref{table sign}.
%%%%%%%%%%%%%%%%%%%%%%%%%%%%%%%%%%
%%%%%%%%%%%%%%%%%%%%%%%%%%%%%%%%%%
\begin{table}[h]
%[htbp]
\begin{center}
\scalebox{0.9}{
\begin{tabular}{|c||c|c|c|c|c|c|c|c|}
\hline
% $(s_L,s_R)$ 
$\ep(s_L,s_R)$
 & $1$ & $-1$& $(-1)^{F_L}$ & $-(-1)^{F_L}$ &  $(-1)^{F_R}$ & $- (-1)^{F_R}$ & $(-1)^{F_L+F_R}$ & $-(-1)^{F_L+F_R}$ \\ \hline\hline
$(\NS,\tNS)$  & $+$ & $-$ & $+$ & $-$ &  $+$ & $-$ & $+$ & $-$ \\ \hline
$(\R,\tNS)$   & $+$ & $-$ & $-$ & $+$ &  $+$ & $-$ & $-$ & $+$  \\ \hline
$(\NS,\R)$    & $+$ & $-$ & $+$ & $-$ &  $-$ & $+$ & $-$ & $+$ \\ \hline
$(\R,\R)$     & $+$ & $-$ & $-$ & $+$ &  $-$ & $+$ & $+$ & $-$ \\ \hline
 \end{tabular}
 }
\caption{Summary of the possible phase choices }
\label{table sign}
\end{center}
\end{table}
%%%%%%%%%%%%%%%%%%%%%%%%%%%%%%%%%%
%%%%%%%%%%%%%%%%%%%%%%%%%%%%%%%%%%

%\begin{description}
%\item[(i)] $\ep^{(s_L,s_R)}_{(1,0), (0,1)} =1$
%\item[(ii)] $\ep^{(s_L,s_R)}_{(1,0), (0,1)} =(-1)^{F_L}$
%\item[(iii)] $\ep^{(s_L,s_R)}_{(1,0), (0,1)} = (-1)^{F_R}$
%\item[(iv)] $\ep^{(s_L,s_R)}_{(1,0), (0,1)} = (-1)^{F_L+F_R}$
%\end{description}
%Then, the phases of remaining sectors $\ep^{(s_L,s_R)}_{(\al,\beta), (a,b)}$ with 
%$(\al,\beta) \neq (0,0)$, $(a,b) \neq (0,0)$ and $(\al,\beta) \neq (a,b)$ are uniquely determined 
%by the modular covariance. 

%%%%%%%%%%%%%%%%%%%%%%%%%%%%%%%%%%%%%%%%%%%%%%%%%%%%%%%%%%%%%%
%%%%%%%%%%%%%%%%%%%%%%%%%%%%%%%%%%%%%%%%%%%%%%%%%%%%%%%%%%%%%%

In the following, we shall separately investigate the vacua of each case;
{\bf (i)} $\ep(s_L,s_R) = 1$,  {\bf (ii)} $\ep(s_L,s_R) = (-1)^{F_L}$, 
{\bf (iii)} $\ep(s_L,s_R) = (-1)^{F_R}$, 
{\bf (iv)} $\ep(s_L,s_R) = (-1)^{F_L+F_R}$.
We then briefly discuss about the remaining four cases, that is, what happens 
if including the overall minus sign into $\ep(s_L,s_R)$.

~

%%%%%%%%%%%%%%%%%%%%%%%%%%%%%%%%%%%%%%%%%%%%%%%%%%%%%%%%%%%
%%%%%%%%%%%%%%%%%%%%%%%%%%%%%%%%%%%%%%%%%%%%%%%%%%%%%%%%%%%
%%%%%%%%%%%%%%%%%%%%%%%%%%%%%%%%%%%%%%%%%%%%%%%%%%%%%%%%%%%

\subsection{Models with the $\cI_4$-orbifolding}

As we discussed above, the models of our interests are classified 
by a single phase factor 
%(depending on the spin structures) 
$\ep(s_L,s_R) \left( \equiv \ep^{(s_L,s_R)}_{(1,0), (0,1)}\right)$.
%$\ep^{(s_L,s_R)}_{(1,0), (0,1)}$, 
%which we simply rewrite as '$\ep$' from now on.
We first focus on the  $\cI_4$-orbifolding.
%%%%

In all the cases, the cylinder amplitudes (self-overlaps) associated to 
D-branes inherited from any BPS bulk-type branes on K3($\cong T^4/ \cI_4$) 
vanish according to the arguments given in \cite{SSU}.

%%%%%%%%%%%%%%%%%%%%%%%%%%%%%%%%%%%%%%%%%%%%%%%%%%%%%%%%%%%%%%%%%%%%

\begin{description}
\item[(1-i) $\ep(s_L,s_R) = 1$ :]

~

This case is just the one analyzed in the previous section, which is the ordinary orbifold defined by the orbifold group 
$\langle \cI_4, \sigma \rangle \cong \bz_2 \times \bz_2 $.
We have the $\cN=(0,1)$ 6-dim. space-time SUSY (8 unbroken supercharges), leading to a vanishing torus partition function; 
$Z^{\msc{torus}}(\tau,\bar{\tau}) =0$. 
All the fractional branes, whose boundary states include  the components of the $\cI_4$-twisted sectors, lead to non-vanishing self-overlap;
$Z^{\msc{cyl}}_{\msc{frac}}(is) \neq 0$. 
%%%%%%%%%%%%%%%%%%%%%%%%%%%%%%%%%%%%%%%%%%%%%%%%%%

%%%%%%%%%%%%%%%%%%%%%%%%%%%%%%%%%%%%%%%%%%%%%%%%%%%%

\item[(1-ii) $\ep(s_L,s_R) = (-1)^{F_L}$ :]

~

In this case the string vacuum has again $\cN=(0,1)$ space-time SUSY.
It is a crucial difference from the first case that the orbifold action is no longer factorized, when taking the spin structures into account. 
This means that the relevant vacua possess non-trivial discrete torsion. 

What happens for the cylinder amplitudes of the fractional branes? 
%As opposed to the first case, 
Due to the extra sign factor $\ep = (-1)^{F_L}$, 
$\sigma$ acts on the $\cI_4$-twisted components of the boundary state as 
\begin{equation}
\sigma \dket{B}_{\cI_4\msc{-twisted}} = (\-_R)^{\otimes 4} \dket{B}_{\cI_4\msc{-twisted}},
\end{equation}
as opposed to the first case. 
We then obtain instead of \eqn{Z-K3frac1}
\begin{align}
\begin{aligned}
	{}_{\msc{frac}}\dbra{B} \sigma\, e^{-\pi s H^{(c)}} \dket{B}_{\text{frac}} &= 
%\cN^2 
\frac{1}{2} \frac{1}{\eta(is)^4} \cdot \left(\frac{2 \eta}{\th_2}\right)^2(is) \cdot  
\frac{1}{\eta(is)^4}\left[ \theta_3^2\theta_4^2  - \theta_4^2\theta_3^2 
%+ \frac{\theta_2(is)^2\theta_1(is)^2}{\eta(is)^4} 
\right](is)\\
	&\hspace{10pt} + 
%\tilde{\cN}^2 
\frac{1}{2}
\frac{1}{\eta(is)^4} \cdot \left(\frac{2 \eta}{\th_3}\right)^2(is) \cdot  \frac{1}{\eta(is)^4} \left[
%\frac{\theta_3(is)^2\theta_1(is)^2}{\eta(is)^4} + 
\theta_4^2\theta_2^2- \theta_2^2\theta_4^2 \right] (is)= 0.
%\\
%	&\sim {\scriptstyle (\bold{1},\sigma)} \underset{(\bold{1},\bold{1})}{\square} + \;{\scriptstyle {(\bold{1},\sigma)}} \underset{(\cI_4,\bold{1})}{\square} \neq 0.
\label{eval Zcyl 1-ii}
\end{aligned}
\end{align}
%%%%%%%%%%%%%%%%
%\begin{align}
%\hspace{-5mm}
%\dbra{B} \sigma e^{-\pi s H^{(c)}} \dket{B}_{\cI^4\msc{-twisted}} & = \frac{1}{\eta(is)^4} \cdot \left(\frac{\eta(is)}{\th_3(is)}\right)^2\cdot 
%\left[\frac{\th_4(is)^2 \th_2(is)^2}{\eta(is)^4} - \frac{\th_2(is)^2 \th_4(is)^2}{\eta(is)^4}\right] \equiv 0.
%\label{eval Zcyl 1-ii}
%\end{align}
%In this evaluation 
The part $[ \cdots ]$ expresses the contribution from world-sheet fermions, of which cancellation implies the bose-fermi degeneracy.

Furthermore, one can examine the massless spectra in the $\cI_4$-twisted sectors for both of the closed and open strings in the same way as the previous section.  
To examine the closed string massless spectra, 
it is convenient to generalize 
%the explicit form of $Z[\cI_4] + Z[\sigma, \cI_4]$ given in 
\eqn{Ztorus 0} to that with the discrete torsion $\ep(s_L,s_R) \equiv \al \beta^{F_L} \gamma^{F_R}$, $(\al, \beta, \gamma =\pm 1)$;
%%%%%
%%%%%
%%%%%
\begin{align}
& 
%\left[ Z[\cI_4] + Z[\sigma, \cI_4] \right]_{\cI_4\msc{-twisted}, \msc{NS-NS}}  
\left[ \bigbox{{\bf 1}}{\cI_4} + \bigbox{\cI_4}{\cI_4}
+ \bigbox{\sigma}{\cI_4} + \bigbox{\sigma\cI_4}{\cI_4} \right]_{\msc{NS-NS}}
= \frac{1}{|\eta|^{16}}
\left|\th_2\right|^8
\left[
\left\{ \left|\th_3\right|^8 + \left|\th_4\right|^8\right\} + \frac{\al}{2} 
\left\{ \th_3^4 \overline{\th_4^4} + \th_4^4 \overline{\th_3^4} \right\}
\right],
\nn
& 
%\left[ Z[\cI_4] + Z[\sigma, \cI_4] \right]_{\cI_4\msc{-twisted}, \msc{R-R}}  
\left[ \bigbox{{\bf 1}}{\cI_4} + \bigbox{\cI_4}{\cI_4}
+ \bigbox{\sigma}{\cI_4} + \bigbox{\sigma\cI_4}{\cI_4} \right]_{\msc{R-R}}
= \frac{1}{|\eta|^{16}}
\left|\th_2\right|^8
\left[
\left\{ \left|\th_3\right|^8 + \left|\th_4\right|^8\right\} - \frac{\al \beta \gamma}{2} 
\left\{ \th_3^4 \overline{\th_4^4} + \th_4^4 \overline{\th_3^4} \right\}
\right].
\label{Ztorus dtorsion}
\end{align}
%%%%%
%%%%%
The present case  corresponds to $\al=1$, $\beta = -1$, $\gamma=1$ and we obtain 
\begin{align}
& 
%\left[ Z[\cI_4] + Z[\sigma, \cI_4] \right]_{\cI_4\msc{-twisted}, \msc{NS-NS}}  = \left[ Z[\cI_4] + Z[\sigma, \cI_4] \right]_{\cI_4\msc{-twisted}, \msc{R-R}} 
\left[ \bigbox{{\bf 1}}{\cI_4} + \bigbox{\cI_4}{\cI_4}
+ \bigbox{\sigma}{\cI_4} + \bigbox{\sigma\cI_4}{\cI_4} \right]_{\msc{NS-NS}}
= \left[ \bigbox{{\bf 1}}{\cI_4} + \bigbox{\cI_4}{\cI_4}
+ \bigbox{\sigma}{\cI_4} + \bigbox{\sigma\cI_4}{\cI_4} \right]_{\msc{R-R}}
\nn
& \hspace{1cm}
= \frac{1}{|\eta|^{16}}
\left|\th_2\right|^8
\left[ \left\{ \left|\th_3\right|^8 + \left|\th_4\right|^8\right\} + \frac{1}{2} 
\left\{ \th_3^4 \overline{\th_4^4} + \th_4^4 \overline{\th_3^4} \right\}\right].
\label{Ztorus 1-ii}
\end{align}
We thus find
$
N^{(\sNS,\sNS)}_{\cI_4\msc{-tw}} = N^{(\sR,\sR)}_{\cI_4\msc{-tw}} = 48.
$
Taking account of the bose-fermi degeneracy, we also conclude
$
N^{(\sNS,\sR)}_{\cI_4\msc{-tw}} = N^{(\sR,\sNS)}_{\cI_4\msc{-tw}} = 48.
$

On the other hand, we can extract the massless open string spectrum in the twisted sector as in \eqn{eval Zcyl 0}. 
As easily seen, the second line of \eqn{eval Zcyl 0} is now replaced with 
$$
 \th_2^4(\th_3^4-\th_4^4)(it) -  \th_2^4(\th_3^4-\th_4^4)(it),
$$
and we find 
$
n^{(\sNS)}_{\msc{tw}} = n^{(\sR)}_{\msc{tw}} =0.
$

~

%%%%%%%%%%%%%%%%%%%%%%%%%%%%%%%%%%%%%%%%%%%%%%%%%%%%%%%%%%%%%%%%%%%%%%%%%

\item[(1-iii) $\ep(s_L,s_R) = (-1)^{F_R}$ :]

~

Since we have
$
(-1)^{F_R }\dket{B}= (-1)^{F_L} \dket{B}
$
for any boundary states, the $\sigma$-orbifolding 
again realizes the vanishing cylinder amplitudes for the fractional branes;
$
Z^{\msc{cyl}}_{\msc{frac}}(is) =0,
$
as in the case {\bf (1-ii)}.

On the other hand, both of the left-moving and right-moving space-time SUSY are broken in the $\cI_4$-twisted sector and we obtain a non-vanishing cosmological constant;
$Z^{\msc{torus}}(\tau,\bar{\tau})\neq 0 $.
%%%%%%%
For instance, 
\begin{align}
\bigbox{\sigma}{\cI_4} + \bigbox{\sigma\cI_4}{\cI_4}& = \frac{1}{\left|\eta\right|^{16}} \frac{1}{2} \left[ \left(\th_3^4 \th_2^4+ \th_2^4 \th_3^4 \right)
\overline{\left(\th_4^4 \th_2^4+ \th_2^4 \th_4^4 \right)} 
+ \left(\th_4^4 \th_2^4+ \th_2^4 \th_4^4 \right)
\overline{\left(\th_3^4 \th_2^4+ \th_2^4 \th_3^4 \right)} 
\right]\neq 0.
\label{eval Z 1-iii}
\end{align}
%%%%
%\begin{align}
%& \hspace{-5mm}
%\tr_{\cI_4\msc{-twsited}} \left[\sigma q^{L_0-\frac{c}{24}}\overline{q^{\tL_0-\frac{c}{24}}} \right] =
%\frac{1}{|\eta(\tau)|^8} \cdot \left(\frac{\eta(\tau)}{\th_4(\tau)}\right)^2 \overline{\left(\frac{\eta(\tau)}{\th_3(\tau)}\right)^2}
%\nn
%& \hspace{1cm}
%\times \left[\frac{\th_3(\tau)^2 \th_2(\tau)^2}{\eta(\tau)^4} + \frac{\th_2(\tau)^2 \th_3(\tau)^2}{\eta(\tau)^4}\right] \cdot 
%\overline{\left[\frac{\th_4(\tau)^2 \th_2(\tau)^2}{\eta(\tau)^4} + \frac{\th_2(\tau)^2 \th_4(\tau)^2}{\eta(\tau)^4}\right]} \neq 0.
%\label{eval Z 1-iii}
%\end{align}
%%%%%%%%%%%%%%%%%%%%%%%

The massless spectra in the $\cI_4$-twisted sector are again examined by 
observing the functions 
%$Z[\cI_4]+ Z[\sigma, \cI_4]$ 
\eqn{Ztorus dtorsion} with $\al=\beta=1$, $\gamma=-1$, 
and $Z_{\msc{frac}}^{\msc{cyl}}$,
and we obtain 
$
N^{(\sNS,\sNS)}_{\cI_4\msc{-tw}} = N^{(\sR,\sR)}_{\cI_4\msc{-tw}} = 48,
$
$
N^{(\sNS,\sR)}_{\cI_4\msc{-tw}} = N^{(\sR,\sNS)}_{\cI_4\msc{-tw}} = 16,
$
$
n^{(\sNS)}_{\msc{tw}} = n^{(\sR)}_{\msc{tw}} =0.
$
Note that the bosonic massless spectrum in the closed string sector is equal with that of the case {\bf (1-ii)}, while the fermionic spectrum is not due to the absence of the bose-fermi cancellation. 

%%%%%%%%%%%%%%%%%%%%%%%%%%%%%%%%%%%%%%%%%%%%%%%%%%%%%%%%%%%
%%%%%%%%%%%%%%%%%%%%%%%%%%%%%%%%%%%%%%%%%%%%%%%%%%%%%%%%%%%
%%%%%%%%%%%%%%%%%%%%%%%%%%%%%%%%%%%%%%%%%%%%%%%%%%%%%%%%%%%

~

\item[(1-iv) $\ep(s_L,s_R) = (-1)^{F_L+F_R}$ :]

~

In this case, we obtain 
$
Z^{\msc{cyl}}_{\msc{frac}}(is) \neq 0,
$
as in the case {\bf (1-i)}, 
since $(-1)^{F_L+ F_R }\dket{B}= (-1)^{F_L} \dket{B}$ holds. 

The right-moving space-time SUSY is broken in the $\cI_4$-twisted sector, 
while the left-moving SUSY is already broken in the untwisted sector ({\em i.e.} with no $\cI_4$-twisting). 
Nevertheless, we find that the cosmological constant vanishes at the one-loop;
$Z^{\msc{torus}}(\tau,\bar{\tau}) =  0 $.
In fact, all the partition functions for the sectors of
$(\al,\beta)=(0,0)$ and $(\al,\beta)=(a,b)$ vanish due to the bose-fermi cancellation in the right-mover, whereas
the $(\al,\beta) \neq (0,0)$ sectors show the left-moving cancellation; 
%%%%%%%%%%%%%%%%%%%%%%%%%%%%%%%%%%%%%%%
\begin{align}
\bigbox{\sigma}{\cI_4} + \bigbox{\sigma\cI_4}{\cI_4}& = \frac{1}{\left|\eta\right|^{16}} \frac{1}{2} \left[ \left(\th_3^4 \th_2^4 - \th_2^4 \th_3^4 \right)
\overline{\left(\th_4^4 \th_2^4+ \th_2^4 \th_4^4 \right)} 
+ \left(\th_4^4 \th_2^4 -  \th_2^4 \th_4^4 \right)
\overline{\left(\th_3^4 \th_2^4+ \th_2^4 \th_3^4 \right)} 
\right] = 0.
\label{eval Z 1-iv}
\end{align}
%%%%%%%%%%%%%%%%%%%%%%%%%%%%%%%%%%%%%%%%
%\begin{align}
%& \hspace{-5mm}
%\tr_{\cI_4\msc{-twsited}} \left[\sigma q^{L_0-\frac{c}{24}}\overline{q^{\tL_0-\frac{c}{24}}} \right] =
%\frac{1}{|\eta(\tau)|^8} \cdot \left(\frac{\eta(\tau)}{\th_4(\tau)}\right)^2 \overline{\left(\frac{\eta(\tau)}{\th_3(\tau)}\right)^2}
%\nn
%& \hspace{1cm}
%\times \left[\frac{\th_3(\tau)^2 \th_2(\tau)^2}{\eta(\tau)^4} - \frac{\th_2(\tau)^2 \th_3(\tau)^2}{\eta(\tau)^4}\right] \cdot 
%\overline{\left[\frac{\th_4(\tau)^2 \th_2(\tau)^2}{\eta(\tau)^4} + \frac{\th_2(\tau)^2 \th_4(\tau)^2}{\eta(\tau)^4}\right]} \equiv 0,
%\label{eval Z 1-iv}
%\end{align}
as opposed to \eqn{eval Z 1-iii}.
%%%%%%%%%%%%%%%%%%%%%%%%%%%%%%%%%%%%%%%%%

The $\cI_4$-twisted massless spectra are found to be 
$
N^{(\sNS,\sNS)}_{\cI_4\msc{-tw}} =  48,
$
$
N^{(\sR,\sR)}_{\cI_4\msc{-tw}} = 16,
$
$
N^{(\sNS,\sR)}_{\cI_4\msc{-tw}} = 16, 
$
$
N^{(\sR,\sNS)}_{\cI_4\msc{-tw}} = 48,
$
$
n^{(\sNS)}_{\msc{tw}} = 16,
$
$
 n^{(\sR)}_{\msc{tw}} =0.
$
Note that we have the bose-fermi cancellation in the left-mover of closed string sector as mentioned above,
and do not in the open string sector.

%%%%%%%%%%%%%%%%%%%%%%%%%
%%%%%%%%%%%%%%%%%%%%%%%%%

\end{description}

%%%%%%%%%%%%%%%%%%%%%%%%%%%%%%%%%%%%%%%%%%%%%%%%%%%%%%%%%%%%
%%%%%%%%%%%%%%%%%%%%%%%%%%%%%%%%%%%%%%%%%%%%%%%%%%%%%%%%%%%%

~

Let us briefly discuss the inclusion of overall minus factor, 
that is, the remaining four cases; 
{\bf (1-i)'} \, $\ep(s_L,s_R) = -1$, ~
{\bf (1-ii)'} \, $\ep(s_L,s_R) = -(-1)^{F_L}$, ~
{\bf (1-iii)'} \, $\ep(s_L,s_R) = -(-1)^{F_R}$, and 
{\bf (1-iv)'} \, $\ep(s_L,s_R) = - (-1)^{F_L+F_R}$.
It is obvious that the bose-fermi degeneracies are still maintained.
However, the massless spectra are changed in the $\cI_4$- ($\tcI_4$-) twisted sector as well as the cylinder amplitudes. 
%%%%%%

As an example, let us pick up the case  {\bf (1-ii)'}.
The relevant parts of torus partition function are given by setting $\al=-1$, $\beta=-1$, $\gamma=1$ in \eqn{Ztorus dtorsion};
\begin{align}
& 
%\left[ Z[\cI_4] + Z[\sigma, \cI_4] \right]_{\cI_4\msc{-twisted}, \msc{NS-NS}}  = \left[ Z[\cI_4] + Z[\sigma, \cI_4] \right]_{\cI_4\msc{-twisted}, \msc{R-R}} 
\left[ \bigbox{{\bf 1}}{\cI_4} + \bigbox{\cI_4}{\cI_4}
+ \bigbox{\sigma}{\cI_4} + \bigbox{\sigma\cI_4}{\cI_4} \right]_{\msc{NS-NS}}
= \left[ \bigbox{{\bf 1}}{\cI_4} + \bigbox{\cI_4}{\cI_4}
+ \bigbox{\sigma}{\cI_4} + \bigbox{\sigma\cI_4}{\cI_4} \right]_{\msc{R-R}}
\nn
& \hspace{1cm}
= \frac{1}{|\eta|^{16}}
\left|\th_2\right|^8
\left[ \left\{ \left|\th_3\right|^8 + \left|\th_4\right|^8\right\} - \frac{1}{2} 
\left\{ \th_3^4 \overline{\th_4^4} + \th_4^4 \overline{\th_3^4} \right\}\right],
\label{Ztorus 1-ii'}
\end{align}
which implies
$
N^{(\sNS,\sNS)}_{\cI_4\msc{-tw}} = N^{(\sR,\sR)}_{\cI_4\msc{-tw}} = N^{(\sNS,\sR)}_{\cI_4\msc{-tw}} = N^{(\sR,\sNS)}_{\cI_4\msc{-tw}} = 16.
$
%%%%%%
The other cases are similarly investigated, and it turns out that the numbers 48 and 16 are just exchanged 
for the spectra of $N^{(s_L,s_R)}_{\cI_4\msc{-tw}}$ ($N^{(s_L,s_R)}_{\tcI_4\msc{-tw}}$) when moving from the `undashed models' to the `dashed ones'.

We can similarly evaluate the cylinder amplitudes of fractional branes;
\begin{align}
Z^{\msc{cyl}}_{\msc{frac}}(is) & \equiv \dbra{B} (1+\sigma)\, e^{-\pi s H^{(c)}} \dket{B}_{\cI^4\msc{-untwisted}} 
+ \dbra{B} (1+\sigma)\, e^{-\pi s H^{(c)}} \dket{B}_{\cI^4\msc{-twisted}} 
\nn
& =
%\frac{1}{4} \frac{1}{\eta(is)^4} \cdot \frac{Z_{\msc{0-modes}}(is)}{\eta(is)^4}\cdot 
%\left\{\left(\frac{\th_3}{\eta}\right)^4 - \left(\frac{\th_4}{\eta}\right)^4 - \left(\frac{\th_2}{\eta}\right)^4\right\} (is)
\frac{1}{2} \frac{1}{\eta(is)^{12}} \cdot \left[ Z_{\msc{zero-modes}}(is)\cdot 
\left(\th_3^4 - \th_4^4 - \th_2^4\right) (is)
+ \left(\th_3^4 \th_4^4- \th_4^4 \th_3^4\right)(is)
\right.
\nn
& \hspace{2cm}
\left. + \left(\th_3^4 \th_2^4- \th_2^4 \th_3^4\right)(is)  \pm  \left(\th_4^4 \th_2^4- \th_2^4 \th_4^4\right) (is) \right] ,
\label{eval Zcyl 1-ii 2}
\end{align}
where the plus and minus in the double signs correspond  respectively to  {\bf (1-ii)} and {\bf (1-ii)'}
and $Z_{\msc{zero-modes}}(is)$ expresses the zero-mode contribution for the bosonic $T^4$-sector.
%%%
One can directly read off the open string spectrum by making the modular transformation $s=1/t$.
Denoting simply $Z_{\msc{zero-modes}}(is) = t^2 \widetilde{Z}_{\msc{zero-modes}}(it)$, 
we obtain 
\begin{align}
Z^{\msc{cyl}}_{\msc{frac}}(i/t) & = \frac{1}{2} \frac{1}{t^2 \eta(it)^{12}}
\cdot \left[ \widetilde{Z}_{\msc{zero-modes}}(it)\cdot 
\left(\th_3^4 - \th_4^4 - \th_2^4\right) (it)
+ \left(\th_3^4 \th_2^4- \th_2^4 \th_3^4\right)(it)
\right.
\nn
& \hspace{2cm}
\left. + \left(\th_3^4 \th_4^4- \th_4^4 \th_3^4\right)(it)  \pm  \left(\th_2^4 \th_4^4- \th_4^4 \th_2^4\right) (it) \right] 
\nn
& \equiv Z^{(\sNS)}(it) + Z^{(\sR)}(it),
\label{eval Zcyl 1-ii 3}
\end{align}
with 
\begin{align}
 & Z^{(\sNS)}(it) \left(= - Z^{(\sR)}(it)\right) =   \frac{1}{2} \frac{1}{t^2 \eta(it)^{12}}
\cdot \left[ \widetilde{Z}_{\msc{zero-modes}}(it)\cdot 
\left(\th_3^4 - \th_4^4 \right) (it) + \left(\th_3^4 \th_2^4 \mp \th_4^4 \th_2^4 \right)(it)\right].
\label{eval Zcyl 1-ii 4}
\end{align}
%%%%%%
Consequently, focusing on the twisted open string sector 
(the second term), 
we find that extra massless excitations appear for the {\bf (1-ii)'}-case, 
%(due to the terms $\sim \th_3^4 \th_2^4(it) + \th_4^4 \th_2^4(it) $ for the $\NS$-sector), 
while they do not for the {\bf (1-ii)}-case as already mentioned.
%the opposite GSO-condition is satisfied for the {\bf (1-ii)} case, 
%while the ordinary GSO-projection is realized for {\bf (1-ii)'}.
%We thus find that extra massless 
Again, the other cases are similarly investigated, and it turns out that the numbers 16 and 0 are exchanged 
for $n^{(s)}_{\msc{tw}}$ when comparing the `undashed' and the `dashed' cases.

~

%%%%%%%%%%%%%%%%%%%%%%%%%%%%%%%%%%%%%%%%%%%%%%%%%%%%%%%%%%%%%%%%%%%%
%%%%%%%%%%%%%%%%%%%%%%%%%%%%%%%%%%%%%%%%%%%%%%%%%%%%%%%%%%%%%%%%%%%%

All these aspects of constructed vacua are summarized in table \ref{table model 1}.

%%%%%%%%%%%%%%%%%%%%%%%%%%%%%%%%%%
%%%%%%%%%%%%%%%%%%%%%%%%%%%%%%%%%%

\begin{table}[h]
%[htbp]
\begin{center}
\scalebox{0.8}{
\begin{tabular}{|c||c|c|c|c|c|c|c|c|}
\hline
name  & {\bf (1-i)} &  {\bf (1-i)'} & {\bf (1-ii)} &  {\bf (1-ii)'} & {\bf (1-iii)} &  {\bf (1-iii)'} & {\bf (1-iv)} &  {\bf (1-iv)'} \\ \hline
$\ep(s_L,s_R)$
 & $1$ & $-1$& $(-1)^{F_L}$ & $-(-1)^{F_L}$ &  $(-1)^{F_R}$ & $- (-1)^{F_R}$ & $(-1)^{F_L+F_R}$ & $-(-1)^{F_L+F_R}$ \\ \hline
$6d$ SUSY  & $(0,1)$ & $(0,1)$ & $(0,1)$ & $(0,1)$ &  $(0,0)$ & $(0,0)$ & $(0,0)$ & $(0,0)$ \\ \hline
$Z^{\msc{torus}}=0$?  & Y & Y & Y & Y &  N & N & Y & Y  \\ \hline
$Z^{\msc{cyl}}_{\msc{frac}}=0$?  & N & N & Y & Y &  Y & Y & N & N \\ \hline
$N_{\cI_4\msc{-tw}}^{(\sNS,\sNS)}$    & $48$ & $16$ & $48$ & $16$ &  $48$ & $16$ & $48$ & $16$ \\ \hline
$N_{\cI_4\msc{-tw}}^{(\sR,\sR)}$      & $16$ & $48$ & $48$ & $16$ &  $48$ & $16$ & $16$ & $48$ \\ \hline
$N_{\cI_4\msc{-tw}}^{(\sNS,\sR)}$     & $48$ & $16$ & $48$ & $16$ &  $16$ & $48$ & $16$ & $48$ \\ \hline
$N_{\cI_4\msc{-tw}}^{(\sR,\sNS)}$     & $16$ & $48$ & $48$ & $16$ &  $16$ & $48$ & $48$ & $16$ \\ \hline
$ n^{(\sNS)}_{\msc{tw}}$     & $16$ & $0$ & $0$ & $16$ &  $0$ & $16$ & $16$ & $0$ \\ \hline
$ n^{(\sR)}_{\msc{tw}}$     & $0$ & $16$ & $0$ & $16$ &  $0$ & $16$ & $0$ & $16$ \\ \hline
 \end{tabular}
}
\caption{Summary of the $\cI_4$-twisted models }
\label{table model 1}
\end{center}
\end{table}

%%%%%%%%%%%%%%%%%%%%%%%%%%%%%%%%%%%%%%%%%%%%%%%%%%%%%%%%%%%%%%%%%%%%%
%%%%%%%%%%%%%%%%%%%%%%%%%%%%%%%%%%%%%%%%%%%%%%%%%%%%%%%%%%%%%%%%%%%%%
%%%%%%%%%%%%%%%%%%%%%%%%%%%%%%%%%%%%%%%%%%%%%%%%%%%%%%%%%%%%%%%%%%%%%

~

\subsection{Models with $\tcI_4$-orbifolding}

We next consider the $\tcI_4 \equiv (-1)^{F_R} \cI_4 $-orbifolding to 
realize the $\mbox{K3} \cong T^4/\bz_2$-sector.
In this case, as opposed to the $\cI_4$-orbifolding, 
the $\sigma \tcI_4$-projection removes all the supercharges in the right-mover. 
We then obtain string vacua with no supercharges.  

%%%%%%%%%%%%%%%%%%%%%%%%%

We recall the fact that the GSO-condition is effectively reversed in the 
$\tcI_4 
%\left(\equiv (-1)^{F_R} \cI_4 \right)
$-twisted sector
(see {\em e.g.} \cite{Sen}, \cite{GSen}), in comparison with that of $\cI_4$. 

%%%%%%%%%%%%%%%%%%%%%%%%%%

To examine the $\tcI_4$-massless spectra, 
it is also convenient to derive the formula  
corresponding to \eqn{Ztorus dtorsion} (we again set $\ep(s_L,s_R) = \al \beta^{F_L} \gamma^{F_R}$, $\al,\beta,\gamma=\pm 1$);
\begin{align}
& 
%\left[ Z[\tcI_4] + Z[\sigma, \tcI_4] \right]_{\tcI_4\msc{-twisted}, \msc{NS-NS}}  
\left[ \bigbox{{\bf 1}}{\tcI_4} + \bigbox{\tcI_4}{\tcI_4}
+ \bigbox{\sigma}{\tcI_4} + \bigbox{\sigma\tcI_4}{\tcI_4} \right]_{\msc{NS-NS}}
= \frac{1}{|\eta|^{16}}
\left|\th_2\right|^8\left\{ \left|\th_3\right|^8 + \left|\th_4\right|^8\right\} - \frac{\al}{2} \frac{1}{|\eta|^{16}}
\left|\th_2\right|^8\left\{ \th_3^4 \overline{\th_4^4} + \th_4^4 \overline{\th_3^4} \right\},
\nn
& 
%\left[ Z[\tcI_4] + Z[\sigma, \tcI_4] \right]_{\tcI_4\msc{-twisted}, \msc{R-R}}  
 \left[ \bigbox{{\bf 1}}{\tcI_4} + \bigbox{\tcI_4}{\tcI_4}
+ \bigbox{\sigma}{\tcI_4} + \bigbox{\sigma\tcI_4}{\tcI_4} \right]_{\msc{R-R}}
= \frac{1}{|\eta|^{16}}
\left|\th_2\right|^8\left\{ \left|\th_3\right|^8 + \left|\th_4\right|^8\right\} - \frac{\al \beta \gamma}{2} \frac{1}{|\eta|^{16}}
\left|\th_2\right|^8\left\{ \th_3^4 \overline{\th_4^4} + \th_4^4 \overline{\th_3^4} \right\}.
\label{Ztorus dtorsion 2}
\end{align}

%%%%%%%%%%%%%%%%%%%%%%%%%%%%%%%%%%%%%%%%%%%%%%%%%%%%%%%%%%%%%%%%%%%%%%%%%%%%%
%%%%%%%%%%%%%%%%%%%%%%%%%%%%%%%%%%%%%%%%%%%%%%%%%%%%%%%%%%%%%%%%%%%%%%%%%%%%%

~

\begin{description}
\item[(2-i) $\ep(s_L,s_R) = 1$ :]

~

In this case, we have $Z^{\msc{torus}}(\tau,\bar{\tau}) \neq 0$, as in ordinary non-SUSY vacua.
In fact, we find 
%%%%
%%%%
\begin{align}
\bigbox{\sigma}{\tcI_4} + \bigbox{\sigma\tcI_4}{\tcI_4}& = \frac{1}{\left|\eta\right|^{16}} \frac{1}{2} \left[ \left(\th_3^4 \th_2^4+ \th_2^4 \th_3^4 \right)
\overline{\left(\th_4^4 \th_2^4+ \th_2^4 \th_4^4 \right)} 
+ \left(\th_4^4 \th_2^4+ \th_2^4 \th_4^4 \right)
\overline{\left(\th_3^4 \th_2^4+ \th_2^4 \th_3^4 \right)} 
\right]\neq 0,
\label{eval Z 2-i}
\end{align}
%%%%
%%%%
%\begin{align}
%& \hspace{-5mm}
%\tr_{\tcI_4\msc{-twsited}} \left[\sigma q^{L_0-\frac{c}{24}}\overline{q^{\tL_0-\frac{c}{24}}} \right] = 
%\frac{1}{|\eta(\tau)|^8} \cdot \left(\frac{\eta(\tau)}{\th_4(\tau)}\right)^2 \overline{\left(\frac{\eta(\tau)}{\th_3(\tau)}\right)^2}
%\nn
%& \hspace{1cm}
%\times \left[\frac{\th_3(\tau)^2 \th_2(\tau)^2}{\eta(\tau)^4} + \frac{\th_2(\tau)^2 \th_3(\tau)^2}{\eta(\tau)^4}\right] \cdot 
%\overline{\left[\frac{\th_4(\tau)^2 \th_2(\tau)^2}{\eta(\tau)^4} + \frac{\th_2(\tau)^2 \th_4(\tau)^2}{\eta(\tau)^4}\right]} \neq 0,
%\label{eval Z 2-i}
%\end{align}
similarly to \eqn{eval Z 1-iii}.

On the other hand, the fractional branes yield vanishing self-overlaps 
in the manner similar to \eqn{eval Zcyl 1-ii}.
%\begin{align}
%\hspace{-5mm}
%\dbra{B} \sigma e^{-\pi s H^{(c)}} \dket{B}_{\tcI^4\msc{-twsited}} & \propto \frac{1}{\eta(is)^4} \cdot \left(\frac{\eta(is)}{\th_3(is)}\right)^2\cdot 
%\left[\frac{\th_4(is)^2 \th_2(is)^2}{\eta(is)^4} - \frac{\th_2(is)^2 \th_4(is)^2}{\eta(is)^4}\right] \equiv 0.
%\label{eval Zcyl 2-i}
%\end{align}

%%%%%%%%%%%%%%%%%%%%%%%%%%%%%%%%%%%%%%%%%%%%%%%%%%%%%%%%%%%%%%%%

By setting $\al=\beta=\gamma=1$ in \eqn{Ztorus dtorsion 2},
the $\tcI_4$-twisted massless spectra are found to be 
$
N^{(\sNS,\sNS)}_{\tcI_4\msc{-tw}} =  N^{(\sR,\sR)}_{\tcI_4\msc{-tw}} = 16,
$
$
N^{(\sNS,\sR)}_{\tcI_4\msc{-tw}} = 
N^{(\sR,\sNS)}_{\tcI_4\msc{-tw}} = 48,
$
and 
$
n^{(\sNS)}_{\msc{tw}} = 
 n^{(\sR)}_{\msc{tw}} =16.
$

%%%%%%%%%%%%%%%%%%%%%%%%%%%%%%%%%%%%%%%%%%%%%%%%%%%%%%%%%%%%%%%%%

~

\item[(2-ii) $\ep(s_L,s_R) = (-1)^{F_L}$ :]

~

In this case, we have $Z^{\msc{torus}}(\tau,\bar{\tau}) \equiv 0$ despite the SUSY breaking. 
In the $\tcI_4$-twisted sector, we find the bose-fermi cancellation in the left-mover as in \eqn{eval Z 1-iii}. 

On the other hand, the fractional branes yield non-vanishing self-overlaps because 
\begin{align}
\hspace{-5mm}
\dbra{B} \sigma e^{-\pi s H^{(c)}} \dket{B}_{\tcI^4\msc{-twsited}} & = \frac{1}{\eta(is)^4} \cdot \left(\frac{2 \eta}{\th_3}\right)^2(is) \cdot 
%\left[\frac{\th_4(is)^2 \th_2(is)^2}{\eta(is)^4} + \frac{\th_2(is)^2 \th_4(is)^2}{\eta(is)^4}\right] 
\frac{1}{\eta(is)^4} \left[\th_4^2 \th_2^4 + \th_2^2 \th_4^2 \right](is)
\neq 0.
\label{eval Zcyl 2-ii}
\end{align}

By setting $\al=1$, $\beta=-1$, $\gamma=1$ in \eqn{Ztorus dtorsion 2}, 
we find 
$
N^{(\sNS,\sNS)}_{\tcI_4\msc{-tw}} =  16,
$
$
N^{(\sR,\sR)}_{\tcI_4\msc{-tw}} = 48,
$
$
N^{(\sNS,\sR)}_{\tcI_4\msc{-tw}} = 48, 
$
$
N^{(\sR,\sNS)}_{\tcI_4\msc{-tw}} = 16,
$
and also
$
n^{(\sNS)}_{\msc{tw}} = 0,
$
$
 n^{(\sR)}_{\msc{tw}} =16.
$

~

%%%%%%%%%%%%%%%%%%%%%%%%%%%%%%%%%%%%%%%%%%%%%%%%%%%%

\item[(2-iii) $\ep(s_L,s_R) = (-1)^{F_R}$ :]

~

In this case, we again obtain $Z^{\msc{torus}}(\tau,\bar{\tau}) \equiv 0$.
However, as opposed to the previous case, the bose-fermi cancellation appears in the right-mover in the $\tcI_4$-twisted sector.
Again, we find $Z^{\msc{cyl}}_{\msc{frac}}(is) \neq 0$, because of the identity $(-1)^{F_R} \dket{B} = (-1)^{F_L} \dket{B}$ for any boundary states.

Similarly to the above cases, 
We  find 
$
N^{(\sNS,\sNS)}_{\tcI_4\msc{-tw}} = 16,
$
$
N^{(\sR,\sR)}_{\tcI_4\msc{-tw}} = 48,
$
$
N^{(\sNS,\sR)}_{\tcI_4\msc{-tw}} = 16, 
$
$
N^{(\sR,\sNS)}_{\tcI_4\msc{-tw}} = 48,
$
$
n^{(\sNS)}_{\msc{tw}} = 0,
$
$
 n^{(\sR)}_{\msc{tw}} =16.
$

~

%%%%%%%%%%%%%%%%%%%%%%%%%%%%%%%%%%%%%%%%%%%%%%%%%%%%%%

\item[(2-iv) $\ep(s_L,s_R) = (-1)^{F_L+F_R}$ :]

~

This case would be the most curious. 
We have $Z^{\msc{torus}}(\tau,\bar{\tau}) \equiv 0$ due to the bose-fermi cancellation in {\em both\/} of the left and right movers despite the absence of unbroken SUSY.   
We also achieve the vanishing self-overlaps for any fractional branes, as in the case {\bf (2-i)}.

We similarly obtain
$
N^{(\sNS,\sNS)}_{\tcI_4\msc{-tw}} =  N^{(\sR,\sR)}_{\tcI_4\msc{-tw}} = 
N^{(\sNS,\sR)}_{\tcI_4\msc{-tw}} = 
N^{(\sR,\sNS)}_{\tcI_4\msc{-tw}} = 48,
$
and 
$
n^{(\sNS)}_{\msc{tw}} = 
 n^{(\sR)}_{\msc{tw}} =0.
$

\end{description}

~

%%%%%%%%%%%%%%%%%%%%%%%%%%%%%%%%%%%%%%%%%%%%%%%%%%%%%%%%%%%%%%%%%%%%%%%%%%%%%%
%These aspects are summarized in \ref{table 2}. 
%%%%%%%%%%%%%
% case 2
%%%%%%%%%%%%%
%\begin{table}[h]
%%[htbp]
%\begin{center}
%\begin{tabular}{|c|c|c|c|c|}
%\hline
% $\ep(s_L,s_R) $  & $1$ & $(-1)^{F_L}$ &  $(-1)^{F_R}$ & $(-1)^{F_L+F_R}$ \\ \hline\hline
%6d SUSY & $(0,0)$ & $(0,0)$ & $(0,0)$ & $(0,0)$ \\ \hline
%$Z^{\msc{torus}} =0 $ ?  & N & Y & Y & Y \\ \hline
%$Z^{\msc{cyl}}_{\msc{frac}} =0$ ? & Y & N & N & Y \\\hline
% \end{tabular}
%\caption{Summary of the Models with $\tcI_4 $-orbifolding }
%\label{table 2}
%\end{center}
%\end{table}
%%%%%%%%%%%%%%%%%%%%%%%%%%%%%%%%%%%%%%%%%%%%%%%%%%%%%%%%%%%%%%%%%%%%

Finally, let us again discuss 
the inclusion of overall minus factor, namely, the `dashed cases'.
%{\bf (2-i)'} \, $\ep(s_L,s_R) = -1$, ~
%{\bf (2-ii)'} \, $\ep(s_L,s_R) = -(-1)^{F_L}$, ~
%{\bf (2-iii)'} \, $\ep(s_L,s_R) = -(-1)^{F_R}$, and 
%{\bf (2-iv)'} \, $\ep(s_L,s_R) = - (-1)^{F_L+F_R}$.
%%%%%%%%%%%%%%%%%%%%%%%%%%%%%%%%%%%%%%%
The relevant arguments are almost parallel, and the aspects of bose-fermi degeneracies remain unchanged.
On the other hand, the GSO-conditions are opposite to the `undashed cases'.
We summarize these results in table \ref{table model 2}.

%In summary, we conclude:
%\begin{itemize}
%\item {\bf (2-i) - (2-iv) : }
%
%In the twisted sectors of closed string, no extra massless states appear both in the left and right movers. 
%On the other hand, in the twisted sectors of open string associated to the fractional branes, we have extra massless states. 
%
%\item {\bf (2-i)' - (2-iv)' : }
%
%In the twisted sectors of closed string, extra massless states appear both in the left and right movers. 
%In the twisted sectors of open string for the fractional branes, we find that the lightest excitations are massive. 
%\end{itemize}

%%%%%%%%%%%%%%%%%%%%%%%%%%%%%%%%%%%%%%%%%%%%%%%%%%%%%%%%%%%%%%%%%%%%%%%%%%%%%%%%%%%%%%%%%%%%%%%
%%%%%%%%%%%%%%%%%%%%%%%%%%%%%%%%%%%%%%%%%%%%%%%%%%%%%%%%%%%%%%%%%%%%%%%%%%%%%%%%%%%%%%%%%%%%%%%

\begin{table}[h]
%[htbp]
\begin{center}
\scalebox{0.8}{
\begin{tabular}{|c||c|c|c|c|c|c|c|c|}
\hline
name  & {\bf (2-i)} &  {\bf (2-i)'} & {\bf (2-ii)} &  {\bf (2-ii)'} & {\bf (2-iii)} &  {\bf (2-iii)'} & {\bf (2-iv)} &  {\bf (2-iv)'} \\ \hline
$\ep(s_L,s_R)$
 & $1$ & $-1$& $(-1)^{F_L}$ & $-(-1)^{F_L}$ &  $(-1)^{F_R}$ & $- (-1)^{F_R}$ & $(-1)^{F_L+F_R}$ & $-(-1)^{F_L+F_R}$ \\ \hline
$6d$ SUSY  & $(0,0)$ & $(0,0)$ & $(0,0)$ & $(0,0)$ &  $(0,0)$ & $(0,0)$ & $(0,0)$ & $(0,0)$ \\ \hline
$Z^{\msc{torus}}=0$?  & N & N & Y & Y &  Y & Y & Y & Y  \\ \hline
$Z^{\msc{cyl}}_{\msc{frac}}=0$?  & Y & Y & N & N &  N & N & Y & Y \\ \hline
$N_{\tcI_4\msc{-tw}}^{(\sNS,\sNS)}$     & $16$ & $48$ & $16$ & $48$ &  $16$ & $48$ & $16$ & $48$ \\ \hline
$N_{\tcI_4\msc{-tw}}^{(\sR,\sR)}$     & $16$ & $48$ & $48$ & $16$ &  $48$ & $16$ & $16$ & $48$ \\ \hline
$N_{\tcI_4\msc{-tw}}^{(\sNS,\sR)}$     & $48$ & $16$ & $48$ & $16$ &  $16$ & $48$ & $16$ & $48$ \\ \hline
$N_{\tcI_4\msc{-tw}}^{(\sR,\sNS)}$     & $48$ & $16$ & $16$ & $48$ &  $48$ & $16$ & $16$ & $48$ \\ \hline
$ n^{(\sNS)}_{\msc{tw}}$     & $16$ & $0$ & $0$ & $16$ &  $0$ & $16$ & $16$ & $0$ \\ \hline
$ n^{(\sR)}_{\msc{tw}}$     & $16$ & $0$ & $16$ & $0$ &  $16$ & $0$ & $16$ & $0$ \\ \hline
 \end{tabular}
}
\caption{Summary of the $\tcI_4$-twisted models }
\label{table model 2}
\end{center}
\end{table}

~

%%%%%%%%%%%%%%%%%%%%%%%%%%%%%%%%%%%%%%%%%%%%%%%%%%%%%%
%%%%%%%%%%%%%%%%%%%%%%%%%%%%%%%%%%%%%%%%%%%%%%%%%%%%%%

\subsection{Massless Spectra in the Other Sectors}

To complete the analysis on the closed string massless spectra,
we next discuss  the other sectors, that is, the untwisted, $\sigma$-twisted, and $\sigma \cI_4$- ($\sigma \tcI_4$-)twisted sectors.
All of them are found to be independent of the choice of the discrete torsion $\ep(s_L,s_R)$.

~

%%%%%%%%%%%%%%%%%%%%%%%%%%%%%%%%%%%

\begin{itemize}
\item {\bf untwisted sector : }

The number of massless states in the untwisted sector just amounts to $\frac{1}{4} $ of that of $T^4$-compactification. 
We thus obtain
\begin{align}
& N_{\msc{untwisted}}^{(\sNS,\sNS)} = N_{\msc{untwisted}}^{(\sR,\sR)} = N_{\msc{untwisted}}^{(\sNS,\sR)} =N_{\msc{untwisted}}^{(\sR,\sNS)} = 16.
\label{massless u}
\end{align}

%%%%%%%%%%%%%%%%%%%%%%%%%%%%%%%%%%%%

\item {\bf $\sigma$-twisted sector : }

The analysis on the $\sigma$-twisted sector is more non-trivial, since the relevant parts of the partition function depend on the discrete torsion, 
which we again parameterize as  
$\ep(s_L,s_R) \equiv \al \beta^{F_L} \gamma^{F_R}$ ($\al,\beta,\gamma \in \pm 1$).
Then, we can evaluate 
\begin{align}
& 
\left[ \bigbox{{\bf 1}}{\sigma} + \bigbox{\sigma}{\sigma}
+ \bigbox{\cI_4}{\sigma} + \bigbox{\sigma\cI_4}{\sigma} \right]_{\msc{NS-NS}}
\nn
& \hspace{0.5cm}
= \frac{1}{|\eta|^{16}} \frac{1}{2} \left[\left\{|\th_3|^8-|\th_4|^8\right\}\left(\th_3^4+\th_4^4\right) \overline{\th_2^4}
+\left|\th_2\right|^8 \left|\th_3^4+\th_4^4\right|^2
+ (\al\gamma) \left(\th_3^4 \th_4^4 +\beta \th_4^4 \th_3^4\right) \overline{\th_2^8}
\right]
\nn
%%%%%
& \hspace{0.5cm}
\equiv \frac{1}{|\eta|^{16}} 
\left[
\frac{1}{4} \left\{\left(\th_3^4+ \th_4^4\right)^2 + 2 (\al \gamma)  \left(\th_3^4 \th_4^4 +\beta \th_4^4 \th_3^4\right) \right\}\overline{\th_2^8}
+ \frac{3}{4} \left|\th_2\right|^8 \left|\th_3^4+\th_4^4\right|^2
\right]
\nn
%%%%%
%%%%%
&
\left[ \bigbox{{\bf 1}}{\sigma} + \bigbox{\sigma}{\sigma}
+ \bigbox{\cI_4}{\sigma} + \bigbox{\sigma\cI_4}{\sigma} \right]_{\msc{R-R}}
%\nn
%& \hspace{0.5cm}
= \frac{1}{|\eta|^{16}} \frac{1}{2} \left[\left|\th_2\right|^8 \left\{|\th_3|^8+|\th_4|^8\right\}
+\left|\th_2\right|^{16} \right] .
\label{Ztorus sigma dtorsion}
\end{align}
%%%%
Note that the R-R amplitude does not depend on the discrete torsion, since \\
$\dsp \left[ \bigbox{\cI_4}{\sigma} + \bigbox{\sigma\cI_4}{\sigma} \right]_{\msc{R-R}}$ trivially vanishes
due to the fermionic zero-modes.
%%%
The massless excitations possess the conformal weights $h_L=h_R=\frac{1}{2}$, and thus correspond to the terms including $\left|\th_2 \right|^8$ in 
\eqn{Ztorus sigma dtorsion}.
Recalling the existence of right-moving bose-fermi degeneracy in this sector, 
we eventually find 
\begin{align}
& N_{\sigma\msc{-tw}}^{(\sNS,\sNS)} = N_{\sigma\msc{-tw}}^{(\sNS,\sR)} = 48,
\hspace{1cm}
 N_{\sigma\msc{-tw}}^{(\sR,\sR)} = N_{\sigma\msc{-tw}}^{(\sR,\sNS)} = 16.
\label{massless sigma-tw}
\end{align} 
%This result is independent of the choice of discrete torsion $\ep(s_L,s_R)$.

%%%%%%%%%%%%%%%%%%%%%%%%%%%%%%%
%We add a small comment: 
%The NS-NS amplitude given in \eqn{Ztorus sigma dtorsion} includes several contributions of $h_L=0$, $h_R=1$.
%They would look tachyonic if focusing on the left mover. 
%%Does it cause a tachyonic instability? 
%However, one has to recall that the level matching condition $h_L-h_R =0$ should be imposed in order to extract the physical spectrum in string theory, 
%while the modular invariance 
%just requires the weaker condition $h_L-h_R \in \bz$.
%In other words, we have to integrate out the modulus of world-sheet torus, and all of the level mismatch terms are removed after
%making the integration of $\tau_1 \equiv \Re \, \tau$
%\footnote
%   {Such 'level mismatch' terms already appear in the asymmetric orbifold $T^4/\sigma$, that is, in the $q$-expansion of 
%$
%\dsp 
%\bigbox{{\bf 1}}{\sigma} + \bigbox{\sigma}{\sigma}.
%$ 
%In the case of \cite{SSW}, on the other hand, 
%$\sigma$ includes the shift operator $\cT_{2\pi R}$, and such terms do not emerge as explicitly shown in \cite{SSW}.
%}.
%In the end, we conclude that the lightest excitations in this sector are massless and their numbers  are counted as \eqn{massless sigma-tw}
%irrespective of the  discrete torsion $\ep(s_L,s_R)$.

%%%%%%%%%%%%%%%%%%%%%%%%%%%%%%%%%%%%%%

\item {\bf $\sigma \cI_4$-twisted sector : }

The relevant parts of partition function are evaluated as 
\begin{align}
& 
\left[ \bigbox{{\bf 1}}{\sigma\cI_4} + \bigbox{\sigma\cI_4}{\sigma\cI_4}
+ \bigbox{\cI_4}{\sigma\cI_4} + \bigbox{\sigma}{\sigma\cI_4} \right]_{\msc{NS-NS}}
\nn
& \hspace{1.5cm}
= \frac{1}{|\eta|^{16}} \frac{1}{2} \left[\left\{|\th_3|^8+|\th_4|^8\right\}\left|\th_2 \right|^8
+\left|\th_2\right|^{16}
+ (\al\beta) \th_2^8 \overline{\left(\gamma \th_3^4 \th_4^4 - \th_4^4 \th_3^4\right)} 
\right],
\nn
%%%%%
&
\left[ \bigbox{{\bf 1}}{\sigma\cI_4} + \bigbox{\sigma\cI_4}{\sigma\cI_4}
+ \bigbox{\cI_4}{\sigma\cI_4} + \bigbox{\sigma}{\sigma\cI_4} \right]_{\msc{R-R}}
%\nn
%& \hspace{0.5cm}
= \frac{1}{|\eta|^{16}} \frac{1}{2} \left[\left|\th_2\right|^8 \left\{|\th_3|^8+|\th_4|^8\right\}
+\left|\th_2\right|^{16} \right] .
\label{Ztorus sigma-cI_4 dtorsion}
\end{align}
%The dependence on the discrete torsion again appears only in the NS-NS sector, and  
The massless spectra are again extracted from the terms including $\left|\th_2\right|^8$, which does not depend on the discrete torsion. 
Taking account of the bose-fermi degeneracy in this sector, we obtain 
\begin{align}
& N_{\sigma\cI_4\msc{-tw}}^{(\sNS,\sNS)} =  N_{\sigma\cI_4\msc{-tw}}^{(\sR,\sR)}  = N_{\sigma\cI_4\msc{-tw}}^{(\sNS,\sR)} =N_{\sigma\cI_4\msc{-tw}}^{(\sR,\sNS)} =  16.
\label{massless sigma cI_4-tw}
\end{align}

%%%%%%%%%%%%%%%%%%%%%%%%%%%%%%%%%%%%%%%%%%%%%%%%%%%%

\item {\bf $\sigma \tcI_4$-twisted sector : }

The analysis of this sector is almost parallel to that of the $\sigma$-twisted sector, if the roles of left and right movers are exchanged. 
We thus obtain the massless spectra
\begin{align}
& N_{\sigma\tcI_4\msc{-tw}}^{(\sNS,\sNS)} = N_{\sigma\tcI_4\msc{-tw}}^{(\sR,\sNS)} = 48,
\hspace{1cm}
 N_{\sigma\tcI_4\msc{-tw}}^{(\sR,\sR)} = N_{\sigma\tcI_4\msc{-tw}}^{(\sNS,\sR)} = 16,
\label{massless sigma tcI_4-tw}
\end{align} 
which is again independent of the discrete torsion $\ep(s_L,s_R)$.

\end{itemize}

~

%%%%%%%%%%%%%%%%%%%%%%%%%%%%%%%%%%%%%%%%%%%%%%%%%%%%%%%
%%%%%%%%%%%%%%%%%%%%%%%%%%%%%%%%%%%%%%%%%%%%%%%%%%%%%%%
%%%%%%%%%%%%%%%%%%%%%%%%%%%%%%%%%%%%%%%%%%%%%%%%%%%%%%%

\subsection{Comments on unitarity and absence of the tachyonic instability}

We here briefly discuss about the unitarity and the absence of tachyonic instability 
of the string vacua constructed above.
The torus partition functions in unitary theories should be $q$-expanded so that all of the coefficients are positive integers.  
%%%
This is not necessarily self-evident in various twisted sectors, that is, those defined by the $\cI_4$ (or $\tcI_4$), $\sigma$, $\sigma \cI_4$ (or $\sigma \tcI_4$) twistings.
Indeed, we would face up to the opposite GSO-projections in the chiral NS-sectors due to the $(-1)^{F_L}$ or $(-1)^{F_R}$-twisting 
for some choices of the discrete torsion $\ep(s_L,s_R)$.
This would potentially give rise to tachyonic excitations, and even worse, the $q$-expansions with negative coefficients.
%that would spoil the unitarity of physical spectrum. 

~

%%%%%%%%%%%%%%%%%%%%%%%%%%%%%%%%%%%%%

\begin{itemize}

\item {\bf $\cI_4$- (or $\tcI_4$-)twisted sector :  }

As was explicitly analyzed above, the lightest excitations are found to be  massless and we do not have any tachyonic states in all the cases. 
Let us focus on the $q$-expansion of the trace; 
%\begin{align}
%&
$ \dsp \left[ \bigbox{{\bf 1}}{\cI_4} + \bigbox{\sigma}{\cI_4} +  \bigbox{\cI_4}{\cI_4} + \bigbox{\sigma\cI_4}{\cI_4} \right]_{\sNS\msc{-}\sNS}$.
%\label{ZNSNS cI4-tw}
%\end{align}
Since the amplitudes 
$
\bigbox{\sigma}{\cI_4},
$
$
\bigbox{\sigma\cI_4}{\cI_4} 
$
belong to the modular orbit $Z[\sigma, \cI_4]$, they
non-trivially depend on the discrete torsion $\ep(s_L,s_R)$.
In any case, all the terms appearing in this function include $\th_2^4$-factors in {\em both\/} of the left- and right-movers, 
which implies $h_L, h_R \geq \frac{1}{2}$ for the vacuum states. 
In this way we have no tachyonic states irrespective of the choice of discrete torsion, in other words, in all the cases {\bf (1-i)}, {\bf (1-i)'}, $\ldots$, {\bf (2-iv)'}.
Furthermore it is just a straightforward task to confirm that it has suitable $q$-expansions whose coefficients are always positive integers 
in all the cases. 

The same arguments are applicable to the other spin structures: R-R, NS-R, R-NS sectors.

%%%%%

\item {\bf $\sigma$ and $\sigma \tcI_4$-twisted sectors :  }

These cases are subtler than the $\cI_4$- ($\tcI_4$-) twisted sector. 
For example, in order to analyze the NS-NS spectrum of the $\sigma$-twisted sector,  
we have to examine the $q$-expansions of torus amplitude \eqn{Ztorus sigma dtorsion}.
%%%
This time, we have the $\th_2^4$-factor only in the right-mover, and the terms $\bigbox{\cI_4}{\sigma}$, $\bigbox{\sigma\cI_4}{\sigma}$ 
($\bigbox{\tcI_4}{\sigma}$, $\bigbox{\sigma\tcI_4}{\sigma}$)
depend on the discrete torsion $\ep(s_L, s_R)$. 
Then, we can confirm that the coefficients of $q$-expansions 
are always positive integers regardless of $\ep(s_L,s_R)$ ($\al$, $\beta$, $\gamma$, in other words).

One would be aware of the existence of several terms with 
$h_L=0$, $h_R=1$ in \eqn{Ztorus sigma dtorsion}, 
and be afraid that tachyonic modes might emerge in the left-mover. 
%%%%
%The NS-NS amplitude given in \eqn{Ztorus sigma dtorsion} includes several contributions of $h_L=0$, $h_R=1$.
%They would look tachyonic if focusing on the left mover. 
%%Does it cause a tachyonic instability? 
However, one has to recall that the level matching condition $h_L-h_R =0$ should be imposed in order to extract the physical spectrum in string theory, 
while the modular invariance 
just requires the weaker condition $h_L-h_R \in \bz$.
In other words, we have to integrate out the modulus of world-sheet torus, and all of the level mismatch terms are removed after
performing the integration of $\tau_1 \equiv \Re \, \tau$
\footnote
   {Such `level mismatch' terms already appear in the asymmetric orbifold $T^4/\sigma$, that is, in the $q$-expansion of 
$
\dsp 
\bigbox{{\bf 1}}{\sigma} + \bigbox{\sigma}{\sigma}.
$ 
In the case of \cite{SSW}, on the other hand, 
$\sigma$ includes the shift operator $\cT_{2\pi R}$, and such terms do not emerge as explicitly shown in \cite{SSW}.
}.
In the end, the lightest excitations in this sector are at most massless and no tachyons emerge in this sector.

We can likewise analyze the $\sigma \tcI_4$-twisted sector.
The coefficients of relevant $q$-expansions are found to be always positive integers irrespective of the discrete torsion. 
We generically have  the terms with $h_L=1$, $h_R=0$, but they do not survive after imposing the level matching condition 
$h_L =h_R$.

%%%%%%%%%%%%%%%%%%%%%%%%%%%%%%%%%%%%%%%%%%%%%%%%%%%%%%%%%%
%%%%%%%%%%%%%%%%%%%%%%%%%%%%%%%%%%%%%%%%%%%%%%%%%%%%%%%%%%
%%%%%%%%%%%%%%%%%%%%%%%%%%%%%%%%%%%%%%%%%%%%%%%%%%%%%%%%%%

\item {\bf $\sigma \cI_4$-twisted sector :  }

Finally let us focus on this remaining case. 
This is similar to the previous one, but there is a crucial difference in the $q$-expansions of the NS-NS (and R-NS) sector.

This time, one observes that the relevant function 
$
\dsp 
\left[ \bigbox{{\bf 1}}{\sigma\cI_4} + \bigbox{\sigma\cI_4}{\sigma\cI_4}
+ \bigbox{\cI_4}{\sigma\cI_4} + \bigbox{\sigma}{\sigma\cI_4} \right]_{\msc{NS-NS}}
$
given in \eqn{Ztorus sigma-cI_4 dtorsion}
includes {\em negative\/} terms in its $q$-expansion, 
when choosing 
$\al\beta =+1$, $\gamma=-1$ (in other words, $\ep= (-1)^{F_R} ~ \mbox{or} ~ - (-1)^{F_L+F_R}$), 
although only positive terms appear otherwise.  
Especially the leading terms with negative coefficients appear at $h_L=1$, $h_R=0$.
%%%
However, it turns out that again all of these `pathological' terms reside  only in the level mismatch sectors, 
and thus do not contribute to the physical string spectrum.

In fact, the relevant term is of the form as 
$
\dsp 
\sim \frac{1}{|\eta|^{16}} \th_2^8 \, \overline{\left[\th_2^8- 4 \th_3^4 \th_4^4 \right]} , 
$
and the (complex conjugate of) right-mover is $q$-expanded as 
\begin{align}
\frac{1}{\eta^8} \left[\th_2^8- 4 \th_3^4 \th_4^4 \right] & = -4 \frac{1}{\eta^8} + \frac{\th_2^8}{\eta^8} + 4 \left[\frac{1}{\eta^8} - \left(\sqrt{\frac{2\eta}{\th_2}}\right)^8\right]
\nn
& = -4 \frac{1}{\eta^8} + \frac{\th_2^8}{\eta^8} + 4 q^{-\frac{1}{3}} \left[\prod_{n=1}^{\infty} \frac{1}{\left(1-q^n\right)^8} 
- \prod_{n=1}^{\infty} \frac{1}{\left(1+q^n\right)^8}\right]
\nn
& = -4 q^{-\frac{1}{3}} + \sum_{n\geq 1} \, a_n q^{n-\frac{1}{3}}, \hspace{1cm} a_n \in \bz_{\geq 0}, ~~ (\any n \geq 1),
\end{align} 
where we made use of the identity $\th_2\th_3\th_4 = 2\eta^3$ in the first line. 
This expansion ensures that  the above statement is correct. 

\end{itemize}

~

%%%%%%%%%%%%%%%%%%%%%%%%%%%%%%%%%%%%%%%%%%%%%%%%%%%%%%%%%%%%%%%%%%
%%%%%%%%%%%%%%%%%%%%%%%%%%%%%%%%%%%%%%%%%%%%%%%%%%%%%%%%%%%%%%%%%%

In this way, we conclude that the torus partition functions are suitably $q$-expanded in the way compatible with unitarity and no tachyonic instability appear for  
all the string vacua we constructed irrespective of the discrete torsion $\ep(s_L,s_R)$.

%%%%%%%%%%%%%%%%%%%%%%%%%%%%%%%%%%%%%%%%%%%%%%%%%%%%%%%%%%%%%%%%%
%%%%%%%%%%%%%%%%%%%%%%%%%%%%%%%%%%%%%%%%%%%%%%%%%%%%%%%%%%%%%%%%%

~

%%%%%%%%%%%%%%%%%%%%%%%%%%%%%%%%%%%%%%%%%%%%%%%%%%%%%%%%%%%%%%%%%%%%%%%%%%%%%%%
%%%%%%%%%%%%%%%%%%%%%%%%%%%%%%%%%%%%%%%%%%%%%%%%%%%%%%%%%%%%%%%%%%%%%%%%%%%%%%%
%%%%%%%%%%%%%%%%%%%%%%%%%%%%%%%%%%%%%%%%%%%%%%%%%%%%%%%%%%%%%%%%%%%%%%%%%%%%%%%

\subsection{Non-BPS fractional branes in the type of \cite{GSen}}

Even though our main purpose in this paper has been achieved in the previous analyses, 
there is another interesting issue. Namely, we would like to focus on the non-BPS fractional branes of the types studied in \cite{GSen} (see also \cite{Sen,BargG}).
As opposed to the above cases, they are already non-BPS {\em before taking the $\sigma$-orbifolding}, and identified with the D-branes wrapped around non-supersymmetric cycles
in K3. The boundary states describing them, which we denote $\dket{B}_{\msc{GS}}$,  contain the NSNS components from the $\cI_4$- ($\tcI_4$-)untwisted sector, 
whereas the RR components from  the $\cI_4$- ($\tcI_4$-)twisted sector with the `ordinary' (`opposite') GSO-projection.
It is notable that these type branes possess the non-vanishing RR-charges and thus could be stable despite their non-BPS properties \cite{BargG}.

%%%%%%%%%%%%%%%%%%%%%%%%%%%%%%%%%%%%%%%%%%%%%%%%%%%%%%%%%%%%%%%%%%%%%%%%%%
Then, the relevant computation of self-overlap of $\dket{B}_{\msc{GS}}$ yields 
\begin{align}
Z^{\msc{cyl}}_{\msc{GS}}(is) & \equiv {}_{\msc{GS}}\dbra{B} e^{-\pi s H^{(c)} } \dket{B}_{\msc{GS}}= \frac{1}{\eta(is)^8} \cdot Z_{\msc{zero-modes}}(is) \cdot 
\frac{1}{2} \frac{1}{\eta(is)^4} 
\left[\th_3^4-\th_4^4\right] (is)
%\left[ \left(\frac{\th_3}{\eta}\right)^4-\left(\frac{\th_4}{\eta}\right)^4\right]  (is) 
\nn
& \hspace{2cm} - \frac{1}{\eta(is)^4} \left(\frac{2\eta}{\th_4} \right)^2(is) \cdot \frac{1}{2} \frac{\th_2^2\th_3^2}{\eta^4}(is)
%\left(\frac{\th_2}{\eta}\right)^2\left(\frac{\th_3}{\eta}\right)^2(is)
\nn
& = \frac{1}{2} \frac{\th_2^4}{\eta^{12}} (is)
%\frac{1}{\eta(is)^8} \left(\frac{\th_2}{\eta}\right)^4 (is) 
\left[ Z_{\msc{zero-modes}}(is) - \th_3(is)^4 \right],
\label{eval GS cyl}
\end{align}
where the first and second terms correspond to the NSNS and RR components, and $Z_{\msc{zero-modes}}(is)$ denotes the zero-mode part of bosonic coordinates along the $T^4$-direction.
For the second equality, we made use of the identities $\th_3^4-\th_4^4 = \th_2^4$ and $2 \eta^3 = \th_2 \th_3 \th_4$. 
%%%
According to the arguments given in \cite{GSen}, one can choose the moduli of $T^4$ as well as the zero-mode part of boundary states $\dket{B}_{\msc{GS}}$ so as to achieve 
the vanishing cylinder amplitude $Z^{\msc{cyl}}_{\msc{GS}}(is) \equiv 0$. Namely, it is enough to choose them so that 
\begin{equation}
Z_{\msc{zero-modes}}(is) = \th_3(is)^4.
\label{GS cond}
\end{equation}
In our set-up, we assumed $T^4  = T^4 [SO(8)]$ for the consistency with the asymmetric orbifolding by $\sigma$, 
and thus this equality can be consistently satisfied.

%%%%%%%%%%%%%%%%%%%%%%%%%%%%%%%%%%%

Now, let us consider the $\sigma$-orbifolding. 
We again define the GS-type fractional brane in the $\sigma$-orbifold as follows;
\begin{equation}
\dket{\cB}_{\msc{GS}} := \sqrt{2} \cP_{\sigma} \dket{B}_{\msc{GS}}.
\end{equation}
Then, 
%We shall focus only on the cases of $Z^{\msc{torus}}\equiv 0$, $Z^{\msc{cyl}}_{\msc{frac}} \equiv 0$, 
%that is, {\bf (1-ii)}, {\bf (1-ii)'},  {\bf (2-iv)}, {\bf (2-iv)'}-cases. 
taking account of the insertion of the operator $1+\sigma$,
%projection $\dsp \frac{1+\sigma}{2}$, 
the cylinder partition function 
\eqn{eval GS cyl} should be modified as 
%%%%
\begin{align}
Z^{' \, \msc{cyl}}_{\msc{GS}}(is) & \equiv {}_{\msc{GS}}\dbra{\cB} e^{-\pi s H^{(c)} } \dket{\cB}_{\msc{GS}} = \frac{1}{\eta(is)^8} \cdot Z'_{\msc{zero-modes}}(is) \cdot \frac{1}{2} 
\frac{1}{\eta(is)^4} \left[\th_3^4 - \th_4^4 \right](is)
%\left[ \left(\frac{\th_3}{\eta}\right)^4-\left(\frac{\th_4}{\eta}\right)^4\right]  (is) 
\nn
& \hspace{1.5cm} + \frac{1}{\eta(is)^4} \cdot   \frac{1}{2} \frac{1}{\eta(is)^8}  \left[\th_3^4 \th_4^4 -\th_4^4 \th_3^4 \right](is)
%\left\{ \left(\frac{\th_3}{\eta}\right)^4(is) \left(\frac{\th_4}{\eta}\right)^4(is) -  \left(\frac{\th_4}{\eta}\right)^4(is)  \left(\frac{\th_3}{\eta}\right)^4(is) \right\}
%\nn
%& \hspace{1.5cm} 
- \frac{1}{\eta(is)^4} \cdot \frac{1}{2} \frac{1}{\eta(is)^8} \left[\th_2^4 \th_3^4 \pm \th_2^4 \th_4^4 \right] (is) 
% \frac{1}{2} \left(\frac{\th_2}{\eta}\right)^4 (is) \left\{ \left(\frac{\th_3}{\eta}\right)^4(is) \pm \left(\frac{\th_4}{\eta}\right)^4(is) \right\}
\nn
& = \frac{1}{2} \frac{\th_2^4}{\eta^{12}} (is) \left[ Z'_{\msc{zero-modes}}(is) - \left(\th_3(is)^4 \pm \th_4(is)^4 \right) \right].
\label{eval GS cyl 2}
\end{align}
Here, the double sign depends on how the involution $\sigma$ acts on the RR-components of the boundary state, which lies in the $\cI_4$- ($\tcI_4$-) twisted sector. 
The amplitude \eqn{eval GS cyl 2} does not vanish generically. However, one can suitably choose the zero-mode components of relevant boundary states
as done in \cite{GSen}. 
This aspect is described as follows;
%%%% 
\begin{itemize}
\item {\bf ($n$-ii), ($n$-iii), ($n$-i)', ($n$-iv)'-cases ~ ($n=1,2$) : }

The plus of the double signs in \eqn{eval GS cyl 2} is realized.
Therefore, the vanishing amplitude is achieved if we assume 
\begin{equation}
Z'_{\msc{zero-modes}}(is) =  \th_3(is)^4 + \th_4(is)^4.
\label{GS cond 2}
\end{equation} 
Namely, the bosonic part of boundary state along the $T^4$-direction should be chosen as the
$\sqrt{2} \times $ Ishibashi state for the basic rep. of $SO(8)_1$.

\item {\bf ($n$-ii)', ($n$-iii)', ($n$-i), ($n$-iv)-cases ~ ($n=1,2$) : }

The minus of the double signs in \eqn{eval GS cyl 2} is realized.
Thus, the vanishing amplitude is again realized if taking 
\begin{equation}
Z'_{\msc{zero-modes}}(is) =  \th_3(is)^4 - \th_4(is)^4.
\label{GS cond 3}
\end{equation} 
However, this assumption is not compatible with the unitarity of open string spectrum, since we have
\begin{equation}
Z'_{\msc{zero-modes}}(i/t) =  t^2 \th_3(it)^4 - t^2 \th_2(it)^4,
\end{equation}
and the negative sign in the second term cannot be consistent with the unitarity in open string Hilbert space.

\end{itemize}

~

%%%%%%%%%%%%%%%%%%%%%%%%%%%%%%%%%%%%%%%%%%%%%%%%%%%%%%%%
%%%%%%%%%%%%%%%%%%%%%%%%%%%%%%%%%%%%%%%%%%%%%%%%%%%%%%%%
%%%%%%%%%%%%%%%%%%%%%%%%%%%%%%%%%%%%%%%%%%%%%%%%%%%%%%%%
%%%%%%%%%%%%%%%%%%%%%%%%%%%%%%%%%%%%%%%%%%%%%%%%%%%%%%%%

\section{Conclusions and discussions}

In this paper, we have studied non-BPS D-branes in the type II string vacua based on asymmetric orbifolds of 
$\mbox{K3} \cong T^4/\bz_2$ as a succeeding work of \cite{SSU}, focusing on the fractional D-branes that contains the contributions from the 
twisted sector of the $\bz_2$-orbifolding. 
We have seen that the cylinder partition functions for these fractional branes do not vanish as opposed to the bulk-type branes studied in \cite{SSU}.   
Nextly, as a main analysis, we studied the extensions of models by including the discrete torsion
{\em depending on the spin structures\/}, which was a crucial point in this paper. 
We classified the relevant string vacua with this torsion, and  
found the vacua that make it possible to gain the vanishing self-overlaps  
for arbitrary fractional non-BPS branes, with keeping the vanishing torus partition function intact. 
Namely, in the classification of vacua given in section \ref{sec dtorsion}, 
we have shown that 
the models {\bf (1-ii)}, {\bf (1-ii)'} (chiral SUSY vacua), and 
 {\bf (2-iv)}, {\bf (2-iv)'} (no SUSY vacua) possess the expected properties. 
(See the tables \ref{table model 1}, \ref{table model 2}.)

We have also analyzed the massless spectra in the twisted sectors and their compatibility with unitarity 
of each model, which non-trivially depend on the discrete torsion. 
In some cases, the GSO projection acts on the twisted sector with the `wrong' sign and 
the relevant $q$-expansions include  negative coefficients that might spoil the unitarity. 
However, it turned out that the negative terms only appear in the sectors with level mismatching 
in our case.  We thus conclude that all the sting vacua constructed above 
would be consistent with unitarity. 
%%%
It is a curious issue to what extent such a feature of unitarity can be generic in asymmetric orbifolds. 
Indeed, the discrete torsion of the type adopted in this paper would offer a novel and simple way of breaking the space-time SUSY with keeping the bose-fermi degeneracies 
both in the closed and open string one-loop amplitudes, 
even though the toroidal models worked out here still remain toy models. 
Therefore, it would be  interesting  to further clarify the universal features of such vacua and to explore possible generalizations beyond the toroidal models. 

%%%%%%%%%%%%%%%%%%%%%%%%%%%%%%%%%%%%%%%%%%%%%%%%%%%%%%%%%%%%%%%%%%%%%%%

Since we start with the type II string on the asymmetric orbifold of 
$\mbox{K3} \cong T^4/\cI_4$, it is also an interesting question to ask what the heterotic dual should be. 
Of course, this is an intriguing  and challenging issue because of the existence of 
discrete torsion among the orbifold actions by $\cI_4$, $\sigma$ and also the GSO-twisting,
which non-perturbatively acts on the Hilbert space of the heterotic side. 
It would be interesting to discuss about the comparisons or possible relations with various heterotic string vacua 
with exponentially suppressed cosmological constants studied {\em e.g.} in 
\cite{Blaszczyk:2014qoa,Angelantonj:2014dia,Faraggi:2014eoa,Abel:2015oxa,
Kounnas:2015yrc,Abel:2017rch}.

%%%%%%%%%%%%%%%%%%%%%%%%%%%%%%%%%%%%%%%%%%%%%%%%%%%%%%%%%%%%%%%%%%%%%%%%
%%%%%%%%%%%%%%%%%%%%%%%%%%%%%%%%%%%%%%%%%%%%%%%%%%%%%%%%%%%%%%%%%%%%%%%%

~

\section*{Acknowledgements}

We would like to thank Yuji Satoh for valuable discussions.

%%%%%%%%%%%%%%%%%%%%%%%%%%%%%%%%%%%%%%%%%%%%%%%%%%%%%%%%%%%%%%%%%%%%%%%%%%%
%%%%%%%%%%%%%%%%%%%%%%%%%%%%%%%%%%%%%%%%%%%%%%%%%%%%%%%%%%%%%%%%%%%%%%%%%%%
%%%%%%%%%%%%%%%%%%%%%%%%%%%%%%%%%%%%%%%%%%%%%%%%%%%%%%%%%%%%%%%%%%%%%%%%%%%
%%%%%%%%%%%%%%%%%%%%%%%%%%%%%%%%%%%%%%%%%%%%%%%%%%%%%%%%%%%%%%%%%%%%%%%%%%%

\newpage

%\vspace{3cm}

\noindent
{\bf \Huge Appendices}

\appendix

\section{Theta functions}

We summarize the conventions of theta functions. 
%%%
\begin{align}
 & \displaystyle \theta_1(\tau,z):=i\sum_{n=-\infty}^{\infty}(-1)^n q^{(n-1/2)^2/2} y^{n-1/2}
  \equiv  2 \sin(\pi z)q^{1/8}\prod_{m=1}^{\infty}
    (1-q^m)(1-yq^m)(1-y^{-1}q^m),
   & \\
 & \displaystyle \theta_2(\tau,z):=\sum_{n=-\infty}^{\infty} q^{(n-1/2)^2/2} y^{n-1/2}
  \equiv 2 \cos(\pi z)q^{1/8}\prod_{m=1}^{\infty}
    (1-q^m)(1+yq^m)(1+y^{-1}q^m), \\
 & \displaystyle \theta_3(\tau,z):=\sum_{n=-\infty}^{\infty} q^{n^2/2} y^{n}
  \equiv \prod_{m=1}^{\infty}
    (1-q^m)(1+yq^{m-1/2})(1+y^{-1}q^{m-1/2}),  
\\
 &  \displaystyle \theta_4(\tau,z):=\sum_{n=-\infty}^{\infty}(-1)^n q^{n^2/2} y^{n}
  \equiv \prod_{m=1}^{\infty}
    (1-q^m)(1-yq^{m-1/2})(1-y^{-1}q^{m-1/2}) . 
\\
%& \Theta_{m,k}(\tau,z):=\sum_{n=-\infty}^{\infty}
% q^{k(n+\frac{m}{2k})^2}y^{k(n+\frac{m}{2k})} ,
%\\
&
\eta(\tau) := q^{1/24}\prod_{n=1}^{\infty}(1-q^n).
\end{align}
 Here, we have set $q:= e^{2\pi i \tau}$, $y:=e^{2\pi i z}$  
 (${}^{\forall} \tau \in {\Bbb H}^+$, ${}^{\forall} z \in {\Bbb C}$),
 and used abbreviations, $\theta_i (\tau) \equiv \theta_i(\tau, 0)$
 ($\theta_1(\tau)\equiv 0$).
%$\Theta_{m,k}(\tau) \equiv \Theta_{m,k}(\tau,0)$.
%

In the main text, we repeatedly make use of the 'Euler identity'
\begin{equation}
\th_3(\tau) \th_4(\tau) \th_2(\tau) = 2\eta(\tau)^3.
\end{equation}

~

%%%%%%%%%%%%%%%%%%%%%%%%%%%%%%%%%%%%%%%%%%%%%%%%%%%%%%
%%%%%%%%%%%%%%%%%%%%%%%%%%%%%%%%%%%%%%%%%%%%%%%%%%%%%%
%%%%%%%%%%%%%%%%%%%%%%%%%%%%%%%%%%%%%%%%%%%%%%%%%%%%%%

\newpage

\section{Modular Orbits}
\label{app:orbit}

We summarize the explicit form of each modular orbit defined in \eqn{def orbits}. 
Each orbit $Z[*]$ is separately modular invariant as confirmed explicitly. 
%%%%%
\begin{itemize}
\item {\bf $Z[{\bf 1}]$ :}
%%%
\begin{align}
Z[{\bf 1}] & =  
Z^{T^4[SO(8)]} \cJ(\tau) \overline{\cJ(\tau)}
\nn
& \equiv \frac{1}{|\eta|^{16}} \cdot \frac{1}{2} \left\{|\th_3|^8 + |\th_4|^8+ |\th_2|^8\right\} \cdot \left|\th_3^4-\th_4^4-\th_2^4 \right|^2.
\label{Z 1}
%\\
%Z^{T^4[SO(8)]}& \equiv \frac{1}{2} \left\{\left|\frac{\th_3}{\eta}\right|^8 + \left|\frac{\th_4}{\eta}\right|^8+ \left|\frac{\th_2}{\eta}\right|^8\right\} ,
%\nn
%\cJ(\tau) & \equiv \left(\frac{\th_3}{\eta}\right)^4 -\left(\frac{\th_4}{\eta}\right)^4 - \left(\frac{\th_2}{\eta}\right)^4 .
\end{align}

%%%%%%%%%%%%%%%%%%%%%%%%%%%%%%%%%%%%%%%%%%%%%%%%%%%%%%%%%%%%

\item {\bf $Z[\cI_4]$ ($Z[\tcI_4]$) :}
\begin{align}
Z[\cI_4]  \left( \equiv Z[\tcI_4] \right) & = \frac{1}{\left|\eta \right|^{16}}
\left[
\left|\th_3 \th_4\right|^4 
\left|\th^2_3 \th_4^2- \th^2_4 \th_3^2
\right|^2
+ \left|\th_3 \th_2\right|^4 
\left|\th^2_3 \th_2^2- \th^2_2 \th_3^2
\right|^2
\right.
\nn
& \hspace{2cm}
\left.
+ \left|\th_4 \th_2\right|^4 
\left|\th^2_4 \th_2^2- \th^2_2 \th_4^2
\right|^2
\right].
 \label{Z cI4}
\end{align}

%\begin{align}
%Z[\cI_4]  \left( \equiv Z[\tcI_4] \right) & = \frac{1}{\left|\eta \right|^8}
%\left[
%\left|\frac{\th_3 \th_4}{\eta^2}\right|^4 \left|\left(\frac{\th_3}{\eta}\right)^2 \left(\frac{\th_4}{\eta}\right)^2-
% \left(\frac{\th_4}{\eta}\right)^2 \left(\frac{\th_3}{\eta}\right)^2\right|^2
%\right.  
%\nn
%& \left. + \left|\frac{\th_3 \th_2}{\eta^2}\right|^4 \left|\left(\frac{\th_3}{\eta}\right)^2 \left(\frac{\th_2}{\eta}\right)^2-
% \left(\frac{\th_2}{\eta}\right)^2 \left(\frac{\th_3}{\eta}\right)^2\right|^2\right.
%\nn
%& \left. + \left|\frac{\th_4 \th_2}{\eta^2}\right|^4 \left|\left(\frac{\th_4}{\eta}\right)^2 \left(\frac{\th_2}{\eta}\right)^2-
% \left(\frac{\th_2}{\eta}\right)^2 \left(\frac{\th_4}{\eta}\right)^2\right|^2\right].
% \label{Z cI4}
%\end{align}

%%%%%%%%%%%%%%%%%%%%%%%%%%%%%%%%%%%%%%%%%%%%%%%%%%%%%%%%%%%%
\item {\bf $Z[\sigma]$ :}
\begin{align}
Z[\sigma] & =  \frac{1}{\left|\eta \right|^{16}} \cdot \frac{1}{2}
 \left[
\left(\th_3^4 + \th_4^4\right)\left(\th_3^4 - \th_4^4 + \th_2^4 \right) \cdot 
\overline{\th_3^2\th_4^2 \left(\th_3^2\th_4^2 - \th_4^2\th_3^2\right)}
\right.
\nn
& 
\hspace{2cm}
+ \left(\th_3^4 + \th_2^4\right)\left(\th_3^4 + \th_4^4 - \th_2^4 \right) \cdot 
\overline{\th_3^2\th_2^2 \left(\th_3^2\th_2^2 - \th_2^2\th_3^2\right)}
\nn
&
\hspace{2cm}
\left. 
-  \left(\th_4^4 - \th_2^4\right)\left(\th_3^4 + \th_4^4 + \th_2^4 \right) \cdot 
\overline{\th_4^2\th_2^2 \left(\th_4^2\th_2^2 - \th_2^2\th_4^2\right)}
\right].
\label{Z sigma}
\end{align}

%%%%%%%%%%%%%%%%%%%%%%%%%%%%%%%%%%%%%%%%%%%%%%%%%%%%%%%%%%%%%%%%%%%%%%%%%

\item {\bf $Z[\cI_4\sigma]$ ($Z[\tcI_4\sigma]$) :}
\begin{align}
Z[\cI_4 \sigma] & = \frac{1}{\left|\eta \right|^{16}} \cdot \frac{1}{2}
\left[ \th_3^2\th_4^2 \left(\th_3^2\th_4^2 - \th_4^2\th_3^2\right) \cdot \overline{\left(\th_3^4 + \th_4^4\right) \cJ(\tau)}
\right.
\nn
& 
\hspace{2cm}
+ \th_3^2\th_2^2 \left(\th_3^2\th_2^2 - \th_2^2\th_3^2\right) \cdot \overline{\left(\th_3^4 + \th_2^4\right) \cJ(\tau)}
\nn
& 
\hspace{2cm}
\left. 
+ \th_4^2\th_2^2 \left(\th_4^2\th_2^2 - \th_2^2\th_4^2\right) \cdot \overline{\left(\th_4^4 - \th_2^4\right) \cJ(\tau)}
\right],
\label{Z I_4-sigma}
\\
%\end{align}
%%%%%%%%%%%%%%%%%%%%%%%%%%%%%%%%%%%%%%%%%%%%%%%%%%%%%%%%%%%%%%%%%%%%%%%%%
%\begin{align}
Z[\tcI_4 \sigma] & = \frac{1}{\left|\eta \right|^{16}} \cdot \frac{1}{2}
\left[ \th_3^2\th_4^2 \left(\th_3^2\th_4^2 - \th_4^2\th_3^2\right) \cdot \overline{\left(\th_3^4 + \th_4^4\right)\left(\th_3^4 - \th_4^4 + \th_2^4 \right) }
\right.
\nn
& 
\hspace{2cm}
+ \th_3^2\th_2^2 \left(\th_3^2\th_2^2 - \th_2^2\th_3^2\right) \cdot \overline{\left(\th_3^4 + \th_2^4\right) \left(\th_3^4 + \th_4^4 - \th_2^4 \right)}
\nn
& 
\hspace{2cm}
\left. 
- \th_4^2\th_2^2 \left(\th_4^2\th_2^2 - \th_2^2\th_4^2\right) \cdot \overline{\left(\th_4^4 - \th_2^4\right) \left(\th_3^4 + \th_4^4 + \th_2^4 \right)}
\right].
\label{Z tI_4-sigma}
\end{align}

%%%%%%%%%%%%%%%%%%%%%%%%%%%%%%%%%%%%%%%%%%%%%%%%%%%%%%%%%%%%%%%%%%%%%%%%

\item {\bf $Z[\sigma, \cI_4]$ 
%(without the discrete torsion; $\ep=1$) 
:}
(the discrete torsion is encoded as $\ep = \al \beta^{F_L} \gamma^{F_R} $ with $\al, \beta, \gamma = \pm 1$)
\begin{align}
Z[\sigma, \cI_4 ] & = \frac{1}{\left|\eta \right|^{16}} \cdot \frac{\al}{2}
\left[
 \th_3^2\th_2^2 \left(\th_3^2\th_2^2 + \beta \th_2^2\th_3^2\right) \cdot 
 \overline{
\th_4^2\th_2^2 \left(\th_4^2\th_2^2 - \gamma \th_2^2\th_4^2\right)
}
\right.
\nn
& \hspace{1cm}
+  \th_4^2\th_2^2 \left(\th_4^2\th_2^2 + \beta \th_2^2\th_4^2\right) \cdot 
 \overline{
\th_3^2\th_2^2 \left(\th_3^2\th_2^2 - \gamma \th_2^2 \th_3^2\right)
}
\nn
& \hspace{1cm}
- \th_3^2\th_4^2 \left(\th_3^2\th_4^2 + \beta \th_4^2\th_3^2\right) \cdot 
 \overline{
\th_4^2\th_2^2 \left(\gamma \th_4^2\th_2^2 - \th_2^2\th_4^2\right)
}
\nn
& \hspace{1cm}
+  \th_3^2\th_4^2 \left(\beta \th_3^2\th_4^2 + \th_4^2\th_3^2\right) \cdot 
 \overline{
\th_3^2\th_2^2 \left(\gamma \th_3^2\th_2^2 - \th_2^2 \th_3^2\right)
}
\nn
& \hspace{1cm}
+ \th_3^2\th_2^2 \left(\beta \th_3^2\th_2^2 + \th_2^2\th_3^2\right) \cdot 
 \overline{
\th_3^2\th_4^2 \left(\gamma \th_3^2\th_4^2 - \th_4^2\th_3^2\right)
}
\nn
& \hspace{1cm}
\left.
+ \th_4^2\th_2^2 \left(\beta \th_4^2\th_2^2 + \th_2^2\th_4^2\right) \cdot 
 \overline{
\th_3^2\th_4^2 \left(\th_3^2\th_4^2 - \gamma \th_4^2\th_3^2\right)
}
\right].
\label{Z sigma I_4}
\end{align}

%%%%%%%%%%%%%%%%%%%%%%%%%%%%%%%%%%%%%%%%%%%%%%%%%%%%%%%%

%%%%%%%%%%%%%%%%%%%%%%%%%%%%%%%%%%%%%%%%%%%%%%%%%%%%%%%%%%%%%%%%%%%%%%%%

\item {\bf $Z[\sigma, \tcI_4]$ 
%(without the discrete torsion; $\ep=1$) 
:}
(the discrete torsion is encoded as $\ep = \al \beta^{F_L} \gamma^{F_R} $ with $\al, \beta, \gamma = \pm 1$)
\begin{align}
Z[\sigma, \tcI_4 ] & = \frac{1}{\left|\eta \right|^{16}} \cdot \frac{\al}{2}
\left[
 \th_3^2\th_2^2 \left(\th_3^2\th_2^2 + \beta \th_2^2\th_3^2\right) \cdot 
 \overline{
\th_4^2\th_2^2 \left(- \th_4^2\th_2^2 - \gamma \th_2^2\th_4^2\right)
}
\right.
\nn
& \hspace{1cm}
-  \th_4^2\th_2^2 \left(\th_4^2\th_2^2 + \beta \th_2^2\th_4^2\right) \cdot 
 \overline{
\th_3^2\th_2^2 \left(\th_3^2\th_2^2 + \gamma \th_2^2 \th_3^2\right)
}
\nn
& \hspace{1cm}
- \th_3^2\th_4^2 \left(\th_3^2\th_4^2 + \beta \th_4^2\th_3^2\right) \cdot 
 \overline{
\th_4^2\th_2^2 \left(\gamma \th_4^2\th_2^2 + \th_2^2\th_4^2\right)
}
\nn
& \hspace{1cm}
+  \th_3^2\th_4^2 \left(\beta \th_3^2\th_4^2 + \th_4^2\th_3^2\right) \cdot 
 \overline{
\th_3^2\th_2^2 \left(\gamma \th_3^2\th_2^2 + \th_2^2 \th_3^2\right)
}
\nn
& \hspace{1cm}
+ \th_3^2\th_2^2 \left(\beta \th_3^2\th_2^2 + \th_2^2\th_3^2\right) \cdot 
 \overline{
\th_3^2\th_4^2 \left(\gamma \th_3^2\th_4^2 + \th_4^2\th_3^2\right)
}
\nn
& \hspace{1cm}
\left.
- \th_4^2\th_2^2 \left(\beta \th_4^2\th_2^2 + \th_2^2\th_4^2\right) \cdot 
 \overline{
\th_3^2\th_4^2 \left(\th_3^2\th_4^2 + \gamma \th_4^2\th_3^2\right)
}
\right].
\label{Z sigma tI_4}
\end{align}

\end{itemize}

%%%%%%%%%%%%%%%%%%%%%%%%%%%%%%%%%%%%%%%%%%%%%%%%%%%%%%%
%%%%%%%%%%%%%%%%%%%%%%%%%%%%%%%%%%%%%%%%%%%%%%%%%%%%%%%
%%%%%%%%%%%%%%%%%%%%%%%%%%%%%%%%%%%%%%%%%%%%%%%%%%%%%%%

~

%%%%%%%%%%%%%%%%%%%%%%%%%%%%%%%%%%%%%%%%%%%%%%%%%%%%%%%%%%%%%%%%%%%%%%%%%%%
%%%%%%%%%%%%%%%%%%%%%%%%%%%%%%%%%%%%%%%%%%%%%%%%%%%%%%%%%%%%%%%%%%%%%%%%%%%
%%%%%%%%%%%%%%%%%%%%%%%%%%%%%%%%%%%%%%%%%%%%%%%%%%%%%%%%%%%%%%%%%%%%%%%%%%%
%%%%%%%%%%%%%%%%%%%%%%%%%%%%%%%%%%%%%%%%%%%%%%%%%%%%%%%%%%%%%%%%%%%%%%%%%%%

\newpage

\begingroup\raggedright\endgroup

\end{document}